**Spin-Splitting Magnetoresistance in Altermagnetic $RuO_2$ Thin Films**

*Hongyu Chen, Zian Wang, Peixin Qin*, Ziang Meng, Xiaorong Zhou, Xiaoning Wang, Li Liu, Guojian Zhao, Zhiyuan Duan, Tianli Zhang, Jinghua Liu*, Dingfu Shao*, Chengbao Jiang*, and Zhiqi Liu**

H. Chen, P. Qin, Z. Meng, X. Zhou, X. Wang, L. Liu, G. Zhao, Z. Duan, T. Zhang, J. Liu, C. Jiang, Z. Liu
School of Materials Science and Engineering, Beihang University, Beijing 100191, China
E-mail: qinpeixin@buaa.edu.cn; 09077@buaa.edu.cn; jiangcb@buaa.edu.cn; zhiqi@buaa.edu.cn

Z. Wang, D. Shao
Key Laboratory of Materials Physics, Institute of Solid State Physics, HFIPS, Chinese Academy of Sciences, Hefei 230031, China
Science Island Branch of Graduate School, University of Science and Technology of China, Hefei 230026, China
E-mail: dfshao@issp.ac.cn

Funding:
National Natural Science Foundation of China (Grant Nos. 52401300, 52425106, 52121001, 52271235, 12274411, 12241405, 52250418, and 524B2003)
National Key R&D Program of China (Grants Nos. 2022YFA1602700 and 2022YFB3506000)
Basic Research Program of the Chinese Academy of Sciences Based on Major Scientific Infrastructures (Grant No. JZHKYPT-2021-08)
CAS Project for Young Scientists in Basic Research (Grant No. YSBR-084)
Beijing Natural Science Foundation (Grant No. JQ23005)
China National Postdoctoral Program for Innovative Talents (Grant No. BX20230451)
China Postdoctoral Science Foundation (Grant No. 2024M754058)

**The recently discovered altermagnets, featured by the exotic correlation of magnetic exchange interaction and alternating crystal environments, have offered exciting cutting-edge opportunities for spintronics. Nevertheless, the altermagnetism of $RuO_2$, one of the earliest-discovered altermagnets, is currently under intense debate. Here we try to resolve this controversy by demonstrating an altermagnetic spin-splitting magnetoresistance (SSMR) effect that is driven by a spin current associated with the giant nonrelativistic spin splitting of an altermagnet. Compared to the spin Hall magnetoresistance induced by a conventional relativistic spin current, the SSMR is characterized by unusual angular dependence with a phase-shift feature underpinned by the Néel-vector orientation and pronounced temperature dependence caused by its susceptibility to electron scattering. Through systematical investigations on the magnetoresistance of (101)-$RuO_2$/Co bilayers, we disentangle a sizable SSMR and hence unveil a Néel vector along [001] direction. Our work not only demonstrates a simple electric avenue to probing the Néel vector of altermagnets, but also indicates long-range magnetic order in thin films of $RuO_2$.**

H. Chen and Z. Wang contributed equally to this work.

## 1. Introduction

Recent theoretical and experimental studies on some unconventional antiferromagnets have unveiled an exotic magnetic phase dubbed altermagnetism.[1–3] It is worth noting that altermagnetism in fact represents a special case of unconventional magnetism (Pomeranchuk instability) in the even partial wave channels that was proposed by Wu *et al*. in the pioneering works in the 2000s.[4,5] Altermagnets are typically composed of two opposite magnetic sublattices that are connected by crystal rotation symmetries. Such alternating magnetic structures break both $\mathcal{PT}$ and $\mathcal{T}\tau$ symmetry ($\mathcal{P}$, $\mathcal{T}$, and $\tau$ denote the operation of space inversion, time reversal, and half-unit-cell translation, respectively), thus engendering sizable spin splitting with alternating spin polarization in momentum space even in the absence of relativistic spin-orbit coupling (SOC).[1–3] Consequently, altermagnets naturally combine the virtues of both ferromagnets and conventional antiferromagnets for spintronics—they are endowed with various $\mathcal{T}$-symmetry-breaking responses[6–10] and charge-spin interconversion phenomena[11–14] that are ubiquitous in ferromagnets despite their symmetry-enforced vanishing net magnetic moment.

Among these exotic properties, the altermagnetic spin-splitting effect (SSE) has received a surge of interest recently.[11,15–18] Taking a *d*-wave altermagnet [**Figure 1**(a)] as an example, the SSE describes the generation of a net nonrelativistic spin current along [100] by an electric current applied along [010] due to the anisotropic spin-split Fermi surfaces as illustrated in Figure 1(b).[11] Remarkably, such a spin current is highly tunable as the spin angular momentum in reciprocal space is closely related to the Néel vector in real space.[11] For instance, an out-of-plane spin current with unconventional spin polarization can be engendered in an altermagnetic thin film through proper epitaxial/Néel-vector orientation design.[15,17–22] On the other hand, the inverse spin-splitting effect (ISSE)—the reciprocal process of the SSE—can convert a spin current with spin polarization parallel to the Néel vector into a transverse charge current. [18–23] Nevertheless, the SSE and ISSE constitute a novel route to realizing charge-spin interconversion—the core of spintronics—without resort to SOC.

The SSE was first proposed in rutile $RuO_2$, a system widely used to discuss altermagnetism and the associated spin-dependent transport properties.[11,15–26] Nevertheless, its magnetic ground state has been under intense debate recently. Although early neutron diffraction,[27] x-ray scattering,[28] and spin- and angle-resolved photoemission spectroscopy studies[29] have indicated the altermagnetism for $RuO_2$, recent first principles calculations[30] and experiments[31–39] question this conclusion, particularly for bulk $RuO_2$. Moreover, the situation is much more intricate for thin films of $RuO_2$. Latest theoretical studies highlight the fragility of the altermagnetism in $RuO_2$,[30,40–42] suggesting that $RuO_2$ could be near a quantum critical transition and the long-range magnetic order could be stabilized possibly by proper doping, epitaxial strain, and surface/interface symmetry breaking. The remarkable differences between thin-film and bulk $RuO_2$ matetials based on these aspects are elaborated in Note 2, Supporting Information. Such claims are indeed supported by the aforementioned early studies and recent experiments on the $\mathcal{T}$-symmetry breaking observed by angle-resolved photoemission spectroscopy,[43] optical second harmonic generation,[44] x-ray magnetic linear dichroism,[45] optical excitation of spin polarization,[46] the anomalous Hall effect,[47] and tunneling magnetoresistance effect[48] in thin films.

In order to solve this controversy, here we first propose that the SSE can lead to a nonrelativistic magnetoresistance parallel to the celebrated spin Hall magnetoresistance (SMR) of a relativistic origin,[49–51] which we dub spin-splitting magnetoresistance (SSMR). We show that the SSMR can be differentiated from the SMR according to their distinct angular and temperature dependence. We next employ the SSMR to explore the possible altermagnetism in epitaxial $RuO_2$ thin films. It is found that the magnetoresistance of (101)-$RuO_2$/Co bilayers contains comparable contributions from the SSMR and the SMR, with the former suggesting a Néel vector along approximately [001] direction. Our work not only demonstrates an unexplored nonrelativistic magnetotransport property of altermagnets, but also indicates the existence of long-range magnetic order in epitaxial $RuO_2$ thin films.

## 2. Results and Discussions

For simplicity, we consider a bilayer composed of a ferromagnet (FM) and a *d*-wave altermagnet with its Néel vector $\boldsymbol{n}$ // [001] as shown in Figure 1(c) to elaborate on the mechanism and characteristics of the SSMR. For a (101)-oriented altermagnetic thin film, an electric field $E$ along [010] generates a spin current $J^{\boldsymbol{p}} = \sigma^{\boldsymbol{p}} E$ along [100] direction, where $\sigma^{\boldsymbol{p}}$ is the associated spin conductivity with the spin polarization $\boldsymbol{p}$. In the absence of SOC, $J^{\boldsymbol{p}}$ is solely engendered by the SSE and $\boldsymbol{p}$ is collinear to $\boldsymbol{n}$ (Fig. 1(b)). Since [100] direction is oblique to the (101) plane, $J^{\boldsymbol{p}}$ flows toward the ferromagnet on the top and is then reflected by the altermagnet/ferromagnet interface, with the reflection strength modulated by the relative orientation between $\boldsymbol{p}$ and the magnetization direction $\boldsymbol{m}$ of the ferromagnet. When $\boldsymbol{p}$ is perpendicular (parallel) to $\boldsymbol{m}$, the reflection is minimized (maximized) due to the strong (weak) absorption of $J^{\boldsymbol{p}}$ by the FM. The reflected spin current thus has the intensity proportional to $(\boldsymbol{m} \cdot \boldsymbol{n})^2$, which converts into an additional longitudinal charge current $J_{\mathrm{rc}} = \theta_{\mathrm{SSMR}} J^{\boldsymbol{p}} (\boldsymbol{m} \cdot \boldsymbol{n})^2 \propto E(\boldsymbol{m}

$\cdot$ $\boldsymbol{n})^2$. Here, we qualitaively introduce $\theta_{\mathrm{SSMR}}$ as the SSMR efficiency about the conversion of the reflected spin current to $J_{\mathrm{rc}}$. Consequently, with $\boldsymbol{m}$ rotating in a plane perpendicular to $E$, the longitudinal resistance minimizes at a rotation angle $\beta = \beta_0$ wherein $\boldsymbol{m}$ is parallel to $\boldsymbol{n}$. Therefore, in the absence of SOC, $\beta_0$ is exactly the out-of-plane tilting angle $\theta_n$ of $\boldsymbol{n}$.

The underlying physics of the SSMR can be compared with that of the SMR in nonmagnet/ferromagnet heterostructures:[49–54] (1) Both effects are underpinned by the modification of the charge-spin interconversion processes in the spin-current source layer by $\boldsymbol{m}$. (2) The SMR stems from the relativistic spin Hall effect (SHE) and (or) the Rashba-Edelstein effect (REE) while the SSMR is of a nonrelativistic origin and is thus in principle more prominent.

On the other hand, the characteristics of the SSMR are distinct from those of the conventional SMR.[49–54] Taking the bilayer shown in Figure 1(c) as an example: (1) For $E$ applied along [010], the $\beta$-dependent magnetoresistance minimizes at $\beta_0$ ~90° in the SMR whereas it minimizes at $\beta_0 = \theta_n$ in the SSMR. (2) For measurements with varied $E$ directions, $\beta_0$ almost remains a constant of 90° in the conventional SMR, but it significantly changes in the SSMR as imposed by the interfacial reflection strength of $\boldsymbol{J}^{\boldsymbol{p}}$ and the anisotropy of the SSE. Therefore, the SSMR can serve as a simple electric method for characterizing the $\boldsymbol{n}$ of altermagnets. (3) The SSMR typically exhibits enhanced temperature dependence compared to the SMR since the Fermi-surface-driven SSE is more susceptible to electron scattering.[11,15,18,23,55,56] These features allow the disentangling of SSMR and SMR in realistic systems.

We next systematically study the magnetoresistance in (101)-$RuO_2$/Co bilayers to ascertain the possible magnetism of the $RuO_2$ thin films. If $RuO_2$ is altermagnetic, the nonrelativistic SSE would lead to the magnetoresistance in concert with the relativistic SHE and REE (denoted as SHE hereafter for simplicity), *i.e.*, the SSMR would coexist with the SMR. Notably, the contribution from the SSE to the out-of-plane spin-current generation/conversion is much more pronounced when $E$ is applied along [010] than along [$\bar{1}$01] according to Figure 1(b) and our first principles calculations (Note 6, Supporting Information). Moreover, the SSMR would yield a $\beta_0$ ~35° (the out-of-plane tilting angle of the $\boldsymbol{n}$ of the $RuO_2$ thin films) in the former case. On the other hand, if $RuO_2$ is nonmagnetic, only the SMR with $\beta_0$ close to 90° would be detected regardless of the direction of $E$. Consequently, distinct angular- and temperature-dependence of the magnetoresistance for $E$ applied along [010] and [$\bar{1}$01] would unambiguously demonstrates the existence of the SSMR and thus the long-range antiferromagnetic order in $RuO_2$.

The detailed preparation processes of the samples can be found in the Supporting Information. **Figure 2**(b) shows the x-ray diffraction pattern of a thick $RuO_2$ thin film deposited onto a (101)-oriented $TiO_2$ substrate. The sharp diffraction peaks of $RuO_2$ (101) together with the clear Kiessig fringes indicate the high epitaxy quality of the sample. It is worth noting that the bilayers exhibit observable exchange bias after post annealing in a magnetic field and subsequent field cooling [Figure 2(c) and Note 2, Supporting Information], indicating the antiferromagnetic order of the $RuO_2$ thin films. In addition, according to a previous x-ray scattering study,[28] the $\boldsymbol{n}$ of $RuO_2$ thin films lies in a direction nearly parallel to the *c* axis. Therefore, for the (101)-oriented thin films, the in-plane projection of $\boldsymbol{n}$ is larger along $[\bar{1}01]$ than along [010], which is expected to result in prominent anisotropy of the exchange bias measured along different in-plane directions. As shown in Figure 2(c), for $RuO_2$(20 nm)/Co(5 nm) samples with exchange bias established along $RuO_2[\bar{1}01]$ and [010], both the magnetic coercivity field and the effective exchange bias field are much smaller in the latter case, which suggests a close-to-[001] orientation of $\boldsymbol{n}$ and is consistent with the previous study.[28]

Hall bars as shown in Figure 2(a) are patterned to investigate the magnetoresistance of a $RuO_2$(3 nm)/Co(2.5 nm) bilayer. The definitions of the coordinate system and the rotation angle $\beta$ of the applied magnetic field are illustrated in Figure 2(d). The current is applied along the $x$ axis and the applied field is 3 T in the measurements unless otherwise specified. As depicted in Figure 2(e), the longitudinal resistance $R(\beta)$ measured along $RuO_2$[010] at 50 K reaches a minimum at $\beta^* \sim 55°$ following $\Delta R(\beta)/R(0) \propto \sin^2(\beta - \beta^*)$, in sharp contrast to that measured along $RuO_2[\bar{1}01]$ wherein $\beta^* \sim 90°$ and $\Delta R(\beta)/R(0) \propto \cos^2\beta$. Here $\Delta R(\beta) = R(\beta) - R(0)$, and $R(0)$ denotes the longitudinal resistance of the sample at $\beta = 0°$. As the phase-shifted angular dependence is unusual, we perform a series of control experiments and confirm that it is an intrinsic property of the bilayer, presumably induced by the (S)SMR-like mechanisms rather than artifacts due to angular offset, the Co layer, or the $RuO_2$ layer. We also find that magnon excitations play a negligible role in leading to the abnormal $\Delta R(\beta)/R(0)$. A detailed discussion can be found in the Supporting Information (Note 4). According to the (S)SMR scenario, $\beta^*$ describes the out-of-plane tilting angle of the $\boldsymbol{p}$ of the engendered spin current. Therefore, the anisotropic $\Delta R(\beta)/R(0)$ indicates that the $\boldsymbol{p}$ contains sizable out-of-plane $z$-components for $E$ // $RuO_2$[010] but is only composed of in-plane $y$-components for $E$ // $RuO_2[\bar{1}01]$. In addition, we define the difference between the maximum and the minimum of $R(\beta)$ as $\Delta R$ and the magnetoresistance ratio as $\Delta R/R(0)$. Note that the $\Delta R/R(0)$ measured along $RuO_2$[010] and $RuO_2[\bar{1}01]$ are both on the order of $10^{-5}$, at least one order of magnitude smaller than those for heavy metal/FM bilayers.[49,51–54] Therefore, the underlying physics of the unusual $\Delta R(\beta)/R(0)$ in Figure 2(e) entails discreet analyses.

We first ascertain the role of Co. We consider a situation according to the SMR scenario wherein the spin current that underpins the anisotropic behavior of the $\Delta R(\beta)/R(0)$ is generated by the Co layer. Generally, the SMR describes the magnetoresistance induced by the modulation of the spin distribution and the spin-charge interconversion processes in the bilayers. According to the SMR theories,[49–51] the magnitude of such magnetoresistance is determined by (1) the spin Hall angle of the spin-source layer, (2) the spin transport properties of the spin-source layer, and (3) the dissipation of the spin current by the other layer. Note that both the magnitude of the magnetoresistance $\Delta R/R(0)$ and the angular dependence of the magnetoresistance $\Delta R(\beta)/R(0)$ are distinct for the measurements along $RuO_2$[010] and along $RuO_2[\bar{1}01]$ [Figure 2(e)]. We thus examine whether the factors (1–3) can lead to such anisotropy when the spin current is generated by Co.

We first explore the effect of factor (1), the spin Hall conductivity $\theta_{SH}$ of Co. We have confirmed that the Co layer deposited at room temperature in our study is polycrystalline in nature via x-ray diffraction. As a result, the $\theta_{SH}$ of Co, as an averaged property of various tiny crystalline grains with stochastic orientations, is supposed to be isotropic and cannot lead to current-direction-dependent $\Delta R/R(0)$. In addition, Co can engender a spin current with the anomalous Hall effect (AHE). In this case, the $\theta_{SH}$ changes with $\beta$. For the generation of an out-of-plane spin current, $\sigma_{SH}$ reaches a maximum at $\beta = 90°$ and a minimum at $\beta = 0°$. Therefore, the $\beta$-dependent $\sigma_{SH}$ of Co cannot lead to the phase shift of the $\Delta R(\beta)/R(0)$ curves measured along $RuO_2$[010]. We next explore the effect of factor (2), the spin transport properties of Co. Similar to (1), we note that (2) is supposed to be isotropic for polycrystalline Co. Also, although (2) of Co could change with $\beta$, it cannot result in the phase shift of the $\Delta R(\beta)/R(0)$ curves measured along $RuO_2$[010]. We subsequently explore the effect of factor (3), the dissipation (absorption and reflection) of the Co-generated spin current by the antiferromagnetism of $RuO_2$. Since epitaxial $RuO_2$ thin films are utilized in our study, (3) could be anisotropic. On the other hand, (3) is supposed not to vary with $\beta$ because the small magnetic field used in our study cannot notably change the property of a nonmagnet or an antiferromagnet with strong magnetic anisotropy. This is supported by the almost identical $\Delta R(\beta)/R(0)$ curves of $RuO_2$ in different magnetic fields (Figure S5, Supporting Information). We also replace the Co layer in the sample with a 3-nm-thick Pt layer to further exclude this possibility. As the $\theta_{SH}$ of Pt is much larger than that of Co, the substitution would lead to more pronounced magnetoresistance if the observed $\Delta R(\beta)/R(0)$ was related to the $\beta$-dependent (3) of $RuO_2$. Nevertheless, as shown in **Figure 3**(a, b), the $\Delta R(\beta)/R(0)$ is negligibly small compared to that in Figure 2(c), and no phase-shift feature can be found.

We finally point out that the experimentally measured efficiency for spin-current generation is subtle for Co compared to that for Ni, NiFe, *etc*.[57,58] due to its weak SOC. Note that the magnitude of the SMR is proportional to the square of the $\theta_{SH}$ of the spin-current source. This further magnifies the incapability of Co of leading to sizable SMR, as is confirmed with a control sample of Cu(1 nm)/Co(2.5 nm). Cu is a light metal with a $\theta_{SH}$ that we expect to be much smaller than that of $RuO_2$. Therefore, the role of Co in generating $\Delta R(\beta)/R(0)$ in the bilayer would be highlighted in this sample. As displayed in Figure 3(c), only the ferromagnetic anisotropic magnetoresistance (AMR) induced by the geometrical-size effect,[59] *i.e.*, $\Delta R(\beta)/R(0) \propto \sin^2\beta$, can be found. Therefore, the anisotropic behavior of $\Delta R(\beta)/R(0)$ in Figure 2(e) cannot originate from the spin current generated by Co, and the Co layer contributes to the $\Delta R(\beta)/R(0)$ with the ferromagnetic AMR effect. Moreover, as depicted in Figure 2(f), the subtraction of the AMR contribution from Figure 2(e) changes the $\beta^*$ and the $\Delta R/R(0)$ measured along $RuO_2$[010] to ~69° and ~$1.3 \times 10^{-4}$, respectively (see Note 5, Supporting Information for detailed discussions and the AMR subtraction processes).

We subsequently investigate the effect of the spin current produced by the $RuO_2$/Co interface with a control sample of $RuO_2$(3 nm)/Cu(1 nm)/Co(2.5 nm). It is found that the insertion of a thin Cu layer reduces the $\Delta R/R(0)$ (Figure S8, Supporting Information). Note that a similar phenomenon has also been reported in a previous study.[49] This can derive from the deteriorated spin-transport efficiency along the out-of-plane direction due to the formation of an additional interface between $RuO_2$ and Cu. Moreover, the $\Delta R(\beta)/R(0)$ measured along $RuO_2$[010] exhibits a $\beta^*$ of ~75° after subtracting the AMR of Co. Here the $\beta^*$ is slightly larger than that in Figure 2(e), probably due to the distinct antiferromagnetic domain population/distribution of $RuO_2$ in this sample. In addition, the robust existence of the phase-shift feature in the samples with and without Cu insertion clearly demonstrates that interfacial spin-current generation plays a negligible role in leading to the abnormal $\Delta R(\beta)/R(0)$.

We thus conclude that the unusual $\Delta R(\beta)/R(0)$ is related to the spin current produced by the $RuO_2$ layer. As we have discussed, the SSE is in good accordance to the anisotropic spin-current-generation/conversion behavior as entailed by the $\beta^*$ in Figure 2(f). In addition, the SHE of a (101)-oriented (nonmagnetic) thin film with a rutile structure can, in principle, exhibit a similar feature too due to the low symmetry.[60] Nevertheless, we notice that such a relativistic mechanism is incapable of yielding considerable out-of-plane spin polarization (see Note 6,

Supporting Information for detailed discussions). Therefore, we deduce that the remarkable phase-shifted $\Delta R(\beta)/R(0)$ is related to the existence of the SSE. Further support for this inference can be found in the anisotropic temperature dependence of the $\Delta R/R(0)$. As depicted in **Figure 4**(a), the $\Delta R/R(0)$ measured along $RuO_2$[010] after subtracting the AMR increases more rapidly with decreased temperature compared to that measured along $RuO_2[\bar{1}01]$. This is well consistent with the fact that the temperature-susceptible SSE plays a more important role in the out-of-plane spin-current generation for $E$ // $RuO_2$[010] than for $E$ // $RuO_2[\bar{1}01]$.

According to the SSMR scenario illustrated in Figure 1(c, d), the $\beta^*$ is anticipated to be exactly the out-of-plane tilting angle (~35°) of the Néel vector of $RuO_2$ when the SSE dominates the charge-spin interconversion in the bilayer. The notable deviation from this value indicates that sizable SMR coexists with the SSMR in the sample. This is reasonable since the SSE of different magnetic domains in the $RuO_2$ layer can compensate each other due to its $\mathcal{T}$-odd nature, leading to a small net effect (see Note 6, Supporting Information for a detailed discussion). Moreover, we notice that the $\beta^*$ measured along $RuO_2$[010] gradually increases with raised temperature [Figure 4(b)], indicating that the $\boldsymbol{p}$ becomes closer to the $y$ axis. This can also be interpretated by the competition between the coexisting SSE and SHE—the SHE plays a more important role in the charge-spin interconversion at high temperature due to the weakening of the SSE by enhanced electron scattering.

We subsequently try to disentangle the SSMR from the measured $\Delta R(\beta)/R(0)$ (see Note 7, Supporting Information for detailed discussions on its effectiveness). According to the SMR theories,[49–51] for a transparent interface,

$$\Delta R/R(0) = \theta_{\mathrm{SH}}^2 \frac{\lambda_{\mathrm{sf}}}{t_{\mathrm{AF}}} \frac{\tanh\left(\frac{t_{\mathrm{AF}}}{2\lambda_{\mathrm{sf}}}\right)}{1 + \frac{\rho_{\mathrm{AF}} t_{\mathrm{F}}}{\rho_{\mathrm{F}} t_{\mathrm{AF}}}} [1 - \frac{1}{\cosh\left(\frac{t_{\mathrm{AF}}}{\lambda_{\mathrm{sf}}}\right)}] \tag{1}$$

wherein $\theta_{\mathrm{SH}}$ and $\rho_{\mathrm{AF}}$ are the spin Hall angle and resistivity of the $RuO_2$ layer, respectively; $\sigma^y$ and $\sigma^z$ are the components of the spin conductivity tensor for generating spin currents with $y$- and $z$-direction polarization, respectively. Note that

$$\theta_{\mathrm{SH}}^2 = \rho_{\mathrm{AF}}^2 [(\sigma^y)^2 + (\sigma^z)^2] \tag{2}$$

wherein $\sigma^y$ and $\sigma^z$ are the components of the spin conductivity tensor for generating spin currents with $y$- and $z$-direction polarization, respectively. Among these quantities, $\sigma^y$, $\sigma^z$, $\lambda_{\mathrm{sf}}$, $\rho_{\mathrm{AF}}$, and $\rho_{\mathrm{F}}$ vary with temperature $T$, leading to the temperature dependence of $\Delta R/R(0)$. In order to highlight the role of the SSE and the SHE, we rewrite Equation 1 with Equation 2 as

$$\Delta R/R(0) = [(\sigma^y)^2 + (\sigma^z)^2][\rho(T)\mathrm{SF}(T)]^2 \tag{3}$$

wherein $\rho(T)$ and $\mathrm{SF}(T)$ are the components of the right-hand side of Equation 3 that are related to the resistivity and $\lambda_{sf}$, respectively. Since the $\beta^*$ is the out-of-plane tilting angle of the spin polarization, we have

$$\sin(\beta^*)\sqrt{\Delta R/R(0)} = \sigma^y\rho(T)\mathrm{SF}(T) \tag{4}$$

$$\cos(\beta^*)\sqrt{\Delta R/R(0)} = \sigma^z\rho(T)\mathrm{SF}(T) \tag{5}$$

For $E$ applied along $RuO_2$[010], we assume that $\sigma^y$ contains contributions from both the SSE and the SHE ($\sigma^y_{010} = \sigma^y_{\mathrm{SSE},010} + \sigma^y_{\mathrm{SHE},010}$) while $\sigma^z$ is only related to the SSE ($\sigma^z_{010} = \sigma^z_{\mathrm{SSE},010}$). Note that temperature can remarkably influence $\sigma^y_{\mathrm{SSE},010}$ and $\sigma^z_{\mathrm{SSE},010}$ but not $\sigma^y_{\mathrm{SHE},010}$. Therefore, we have

$$\sin(\beta^*)\sqrt{\Delta R/R(0)}_{010} = \sigma^y_{\mathrm{SSE},010}(T)\rho_{010}(T)\mathrm{SF}_{010}(T) + \sigma^y_{\mathrm{SHE},010}\rho_{010}(T)\mathrm{SF}_{010}(T) \tag{6}$$

$$\cos(\beta^*)\sqrt{\Delta R/R(0)}_{010} = \sigma^z_{\mathrm{SSE},010}(T)\rho_{010}(T)\mathrm{SF}_{010}(T) \tag{7}$$

We define

$$\beta_0 = \tan^{-1}\left(\frac{\sigma^y_{\mathrm{SSE}}}{\sigma^z_{\mathrm{SSE}}}\right) \tag{8}$$

According to the SSE scenario, $\beta_0$ is supposed to be the out-of-plane tilting angle of the Néel vector.

On the other hand, for $E$ applied along $RuO_2$[$\bar{1}$01], there is only finite $\sigma^y$ since $\beta^*$ ~90°. Note that the calculated $\sigma^y_{\mathrm{SSE},\bar{1}01}$ is much smaller in this case (even smaller than the $\sigma^y_{\mathrm{SHE},\bar{1}01}$) (Table S1, Supporting Information). Moreover, the multidomain state of $RuO_2$ can further decrease the realistic $\sigma^y_{\mathrm{SSE},\bar{1}01}$. We thus assume a negligible $\sigma^y_{\mathrm{SSE},\bar{1}01}$ compared to $\sigma^y_{\mathrm{SHE},\bar{1}01}$, and we have

$$\sqrt{\Delta R/R(0)}_{\bar{1}01} = \sigma^y_{\mathrm{SHE},\bar{1}01}\rho_{\bar{1}01}(T)\mathrm{SF}_{\bar{1}01}(T) \tag{9}$$

Since we find a subtle difference between the sample resistance measured along $RuO_2$[010] and $RuO_2$[$\bar{1}$01], we have

$$\rho_{010}(T) \approx \rho_{\bar{1}01}(T) \tag{10}$$

Although the values of SF($T$) along $RuO_2$[010] and $RuO_2$[$\bar{1}$01] can be different, its relative change with varied temperature (from 50 to 300 K) is expected to be nearly identical, because the spin diffusion processes are underpinned by the same mechanisms. In addition, the $\sigma^y_{\mathrm{SHE}}$ can be anisotropic along $RuO_2$[010] and $RuO_2$[$\bar{1}$01]. We thus utilize a constant $\eta$ to describe the anisotropy of $\sigma^y_{\mathrm{SHE}}$ and SF($T$) along $RuO_2$[010] and $RuO_2$[$\bar{1}$01].

$$\eta = \frac{\sigma^y_{\mathrm{SHE},010}\mathrm{SF}_{010}(T)}{\sigma^y_{\mathrm{SHE},\bar{1}01}\mathrm{SF}_{\bar{1}01}(T)} \tag{11}$$

Thus, we have

$$\eta\sqrt{\Delta R/R(0)}_{\bar{1}01} = \sigma^y_{\mathrm{SHE},010}\rho_{010}(T)\mathrm{SF}_{010}(T) \tag{12}$$

Taking Equation. 7, 8, and 12 into Equation 6, we have

$$\sin(\beta^*)\sqrt{\Delta R/R(0)}_{010} = \tan(\beta_0)\cos(\beta^*)\sqrt{\Delta R/R(0)}_{010} + \eta\sqrt{\Delta R/R(0)}_{\bar{1}01} \tag{13}$$

Equation 13 is the core of our method, with which we can fit the measured relation between $\sqrt{\Delta R/R(0)}_{010}$, $\sqrt{\Delta R/R(0)}_{\bar{1}01}$, and $\beta^*$ utilizing two parameters, $\beta_0$ and $\eta$. We can check the fitting effectiveness through recalculating $\beta_0$ with $\eta$ according to

$$\beta_0 = \tan^{-1}\left(\frac{\sigma^y_{\mathrm{SSE}}}{\sigma^z_{\mathrm{SSE}}}\right) = \tan^{-1}\left[\frac{\sin(\beta^*)\sqrt{\Delta R/R(0)}_{010} - \eta\sqrt{\Delta R/R(0)}_{\bar{1}01}}{\cos(\beta^*)\sqrt{\Delta R/R(0)}_{010}}\right] \tag{14}$$

In addition, we can determine the relative strength of the pure SSMR with respect to the pure SMR as

$$\frac{(\sigma^y_{\mathrm{SSE},010})^2 + (\sigma^z_{\mathrm{SSE},010})^2}{(\sigma^y_{\mathrm{SHE},010})^2} = \frac{[\tan(\beta_0)\cos(\beta^*)\sqrt{\Delta R/R(0)}_{010}]^2 + [\cos(\beta^*)\sqrt{\Delta R/R(0)}_{010}]^2}{[\eta\sqrt{\Delta R/R(0)}_{\bar{1}01}]^2} \tag{15}$$

The best fitting gives $\beta_0$ ~33.6° and $\eta$ ~0.84. Note that the calculated $\beta_0$ values at different temperatures with the obtained $\eta$ according to Equation 14 are all close to 33.6° [Figure 4(c)], confirming the reliability of the fitting. Moreover, the $\beta_0$ is close to the out-of-plane tilting angle of the Néel vector,[28] which further corroborates the existence of the SSE. Moreover, the pure SSMR is comparable to the pure SMR in the sample [Figure 4(d)], with its relative strength decreasing with raised temperature. This indicates that the SSMR is more sensitive to electron scattering compared to the SMR, also consistent with the Fermi-surface origin of the SSE.

Our magnetoresistance measurements, together with the exchange-bias effect, provide compelling evidence for the altermagnetism in the epitaxial $RuO_2$ thin films. In addition, we note the fact that the SSMR is strongly suppressed at high temperature also offers an implication that high-temperature detections of altermagnetism might be tricky, due to the spin-splitting of Fermi surface being largely smeared out at high temperatures. In this sense, our results could also be beneficial for resolving the current controversy of different measurements on the altermagnetism of $RuO_2$.

## 3. Conclusion

In summary, we propose a nonrelativistic magnetoresistance effect termed the SSMR for altermagnets. The existence of the SSMR is unambiguously demonstrated for (101)-$RuO_2$/Co bilayers, indicating the presence of altermagnetism in epitaxial $RuO_2$ thin films. The unique phenomenology of the SSMR makes it a simple electric probe for detecting the Néel vector of altermagnets without resort to complex magnetic tunnel junctions[61–64] or lock-in techniques.[65–68] Finally, we point out that the SSMR can also exist in other unconventional antiferromagnets with momentum-dependent spin splitting.[69–75] Therefore, our work opens an avenue to study the emerging materials with unconventional spin degeneracy lifting.

## 4. Experimental

*Sample fabrication*.—The $RuO_2$ thin films in this study were fabricated by a pulsed laser deposition system with a base pressure of $1.5 \times 10^{-8}$ Torr and a target-substrate distance of ~60 mm following a recipe developed in our previous work.[7] The laser fluence and repetition rate were set to be ~1.6 J $cm^{-2}$ and 10 Hz, respectively. During deposition, the substrate temperature and the oxygen pressure were 550°C and 1 mTorr, respectively. After deposition, the samples were *in situ* annealed for ~0.5 h in the chamber under the same conditions (temperature and oxygen pressure) as in the deposition process and then cooled down to room temperature at a ramp rate of ~10°C $min^{-1}$. The samples were subsequently transferred into a d.c. magnetron sputtering system with a base pressure of $7.5 \times 10^{-9}$ Torr and a sample-target distance of ~245 mm. The Co and Al layers were grown at room temperature with a sputtering power of 60 and 30 W, respectively. The Ar pressure during deposition was 3 mTorr. The as-deposited thin films were then patterned into Hall bars through optical lithography and Ar ion milling. The channel width of the Hall bars was 3 μm and the distance between two longitudinal voltage probes was 15 μm.

*X-ray diffraction*.—The x-ray diffraction measurement was performed with a Panalytical MRD X'Pert 3 diffractometer using Cu $K\alpha$ radiation.

*Magnetic and electrical measurements*.—The magnetic properties of the samples were examined by a Quantum Design VersaLab system with a vibrating sample magnetometer module. The transport properties of the samples were measured by the Quantum Design VersaLab system with a horizontal rotator module. A Keithley 6221 source meter and a Keithley 2182A nanovoltmeter were additionally employed to detect weak electrical signals. For each set of $\beta^*$ and $\Delta R/R(0)$, we performed at least two independent $\beta$-dependent magnetoresistance measurements under the same conditions with $\beta$ rotated from 0° to 360° and from 360° back to 0°. The averaged $R(\beta)$ curve was employed to extract $\beta^*$ and $\Delta R/R(0)$ according to a $\sin^2(\beta)$ fitting.

**Supporting Information**

Supporting Information is available from the Wiley Online Library or from the author.

**Acknowledgements**

P.Q. acknowledges funding from National Natural Science Foundation of China (Grant No. 52401300). Z.L. acknowledges funding from National Natural Science Foundation of China (Grant No. 52425106). Z.L. and C.J. acknowledge funding from National Natural Science Foundation of China (Grant No. 52121001). Z.L. acknowledges funding from National Natural Science Foundation of China (Grant No. 52271235) and from National Key RandD Program of China (Grants Nos. 2022YFA1602700 and 2022YFB3506000). D.S. acknowledges funding from National Natural Science Foundation of China (Grants Nos. 12274411, 12241405, and 52250418), from Basic Research Program of the Chinese Academy of Sciences Based on Major Scientific Infrastructures (Grant No. JZHKYPT-2021-08), and from CAS Project for Young Scientists in Basic Research (Grant No. YSBR-084). Z.L. acknowledges funding from Beijing Natural Science Foundation (Grant No. JQ23005). P.Q. acknowledges funding from China National Postdoctoral Program for Innovative Talents (Grant No. BX20230451) and from China Postdoctoral Science Foundation (Grant No. 2024M754058). D.S. acknowledges Hefei Advanced Computing Center. This work is also supported by "the Fundamental Research Funds for the Central Universities". Z.M. acknowledges financial support from National Natural Science Foundation of China (Grant No. 524B2003). This work was also supported by the Academic Excellence Foundation for Ph.D. students of Beihang University.

**Data Availability Statement**

The data that support the findings of this study are available from the corresponding author upon reasonable request.

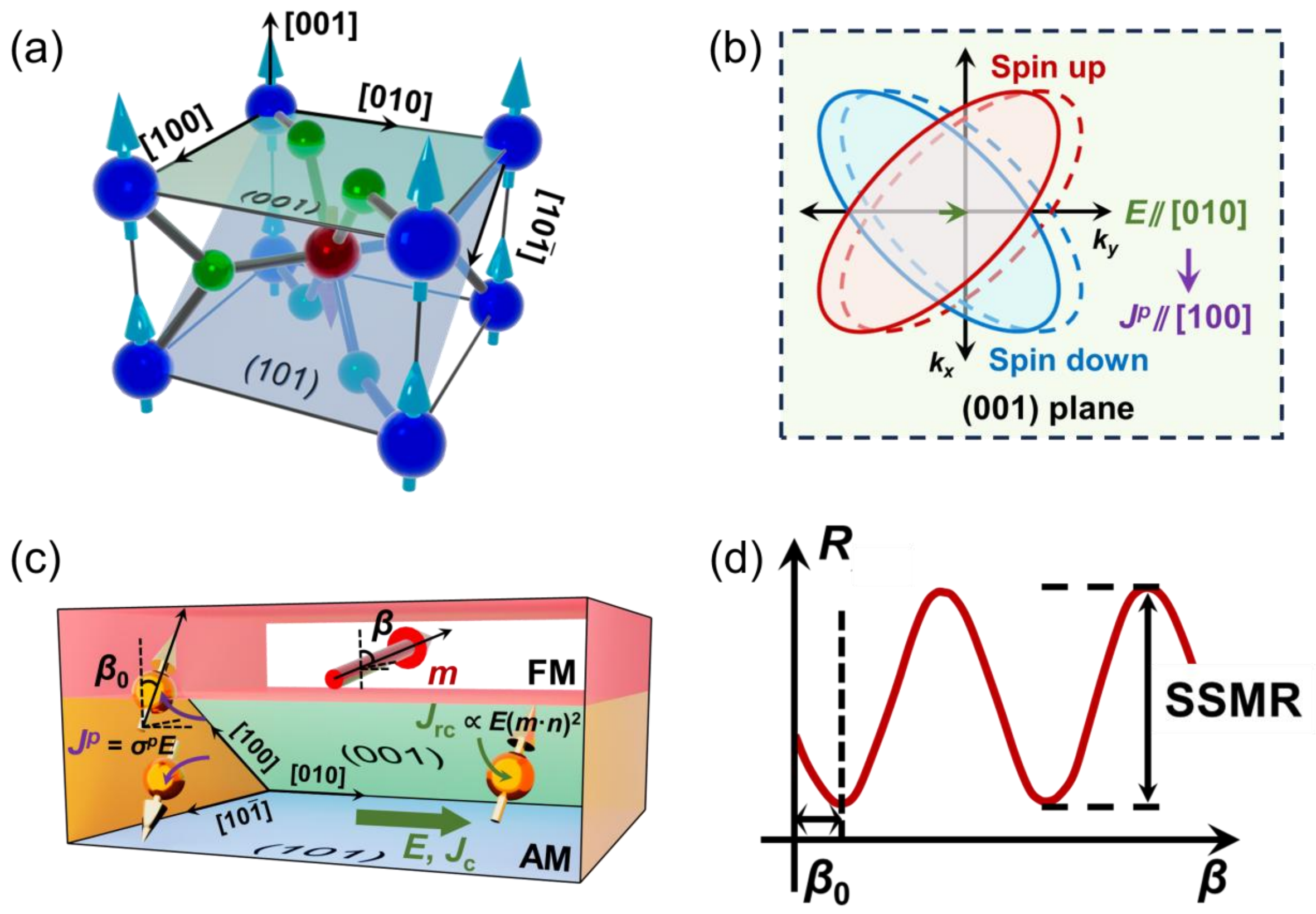

**Figure 1.** (a) Crystal structure of a rutile-structured *d*-wave altermagnet. The green and blue shadowed regions are the (001) and (101) plane, respectively. The dark bule and dark red balls represent the magnetic atoms *M* with opposite magnetic moment as indicated by the light blue and light red arrows, respectively. The green balls denote the nonmagnetic atoms. (b) Schematic of the spin-splitting effect (SSE). An electric field $E$ applied along [010] shifts the anisotropic spin-spilt Fermi contours (the red and blue shadowed ellipses), resulting in a net nonrelativistic spin current $J^p$ along [100] with polarization $\boldsymbol{p}$. $\boldsymbol{p}$ is parallel to the Néel vector $\boldsymbol{n}$. (c) Schematic of the SSMR in a (101)-oriented altermagnet (AM)/ferromagnet (FM) bilayer. An applied $E$ (charge current $J_c$) along [010] engenders an out-of-plane $J^p = \sigma^p E$ through the spin conductivity $\sigma^p$ of the SSE, illustrated by the movement of electrons (oranges balls) with their spin moment (yellow arrows) parallel to $\boldsymbol{n}$. $\beta_0$ is the out-of-plane tilting angle of $\boldsymbol{n}$. $J^p$ is reflected by the AM/FM interface, leading to a longitudinal charge current $J_{cr} \propto E(\boldsymbol{m}\cdot\boldsymbol{n})^2$ that lowers the resistance of the heterostructure. The magnitude of $J_{cr}$ is dictated by the squared product of $\boldsymbol{n}$ and the magnetization unit vector $\boldsymbol{m}$ of the FM. (d) Envisaged longitudinal resistance $R$ as a function of $\beta$ in the SSMR.

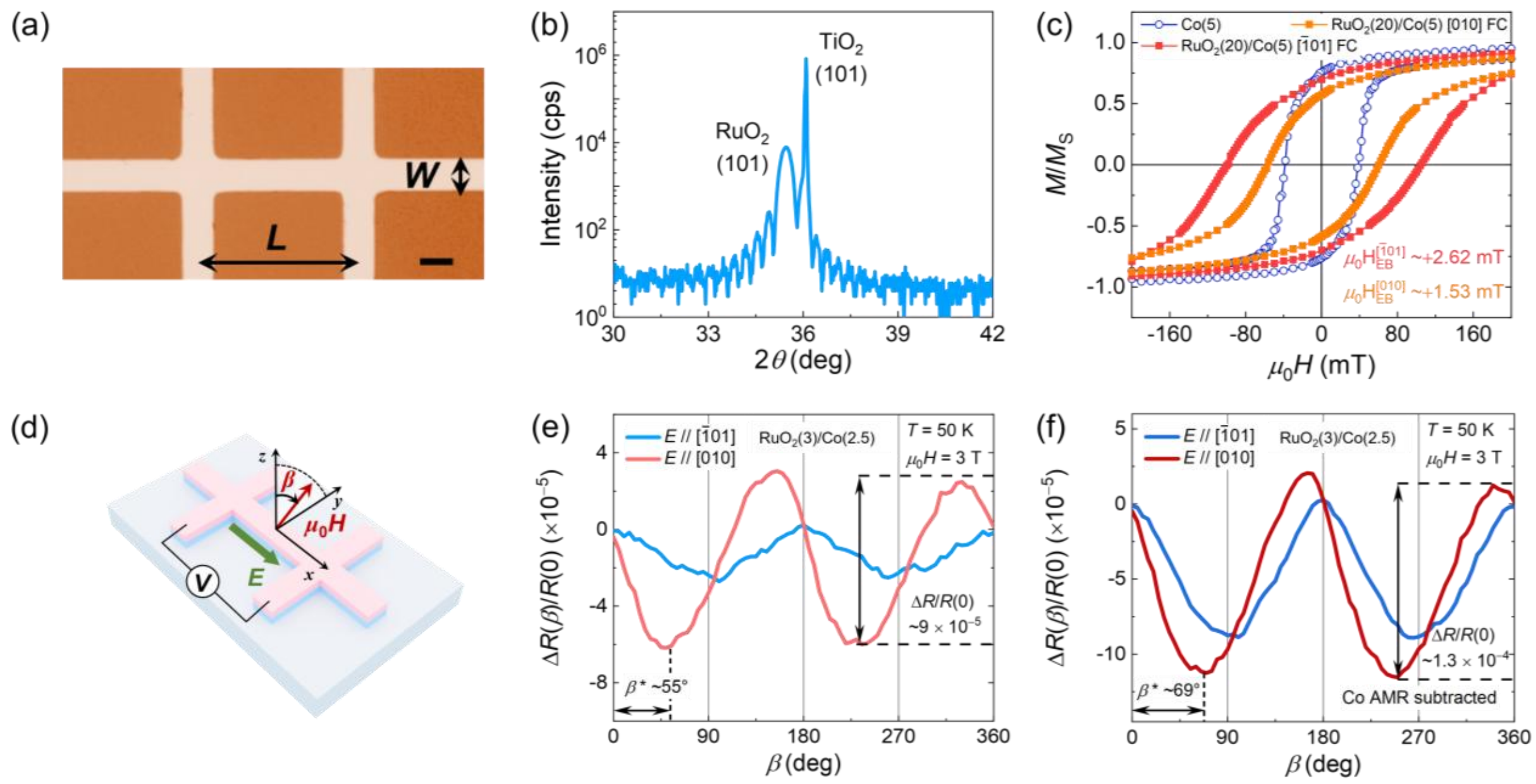

**Figure 2.** (a) Optical image of a typical Hall bar device with width $W$ and length $L$. The scale bar is 3 μm. (b) Enlarged x-ray diffraction pattern of a thick $RuO_2$ sample on a (101)-$TiO_2$ substrate. The full-range pattern can be found in the Supporting Information. (c) Normalized magnetization $M/M_S$ as a function of external magnetic field $\mu_0 H$ at 50 K of a single Co layer and a $RuO_2$(20 nm)/Co(5 nm) bilayer. In order to establish exchange bias, both samples are post-annealed at 393 K for 0.5 h (with a magnetic field of –1.5 T applied along $RuO_2[\bar{1}01]$ and [010]). The measurements are conducted after field cooling (FC) in the same magnetic field. (d) Schematic of the coordinate system and the rotation angle $\beta$ of the magnetic field $\mu_0 H$ in our measurements. The longitudinal voltage $V$ is detected with $E$ applied along the $x$ axis. (e, f) $\beta$-dependent magnetoresistance $\Delta R(\beta)/R(0)$ of the $RuO_2$(3 nm)/Co(2.5 nm) bilayer measured at 50 K along $RuO_2$[010] and $RuO_2[\bar{1}01]$ in a magnetic field of 3 T before and after the subtraction of the contribution from the ferromagnetic anisotropic magnetoresistance (AMR) of Co. $R(0)$ denotes the $R$ of the bilayer at $\beta = 0°$.

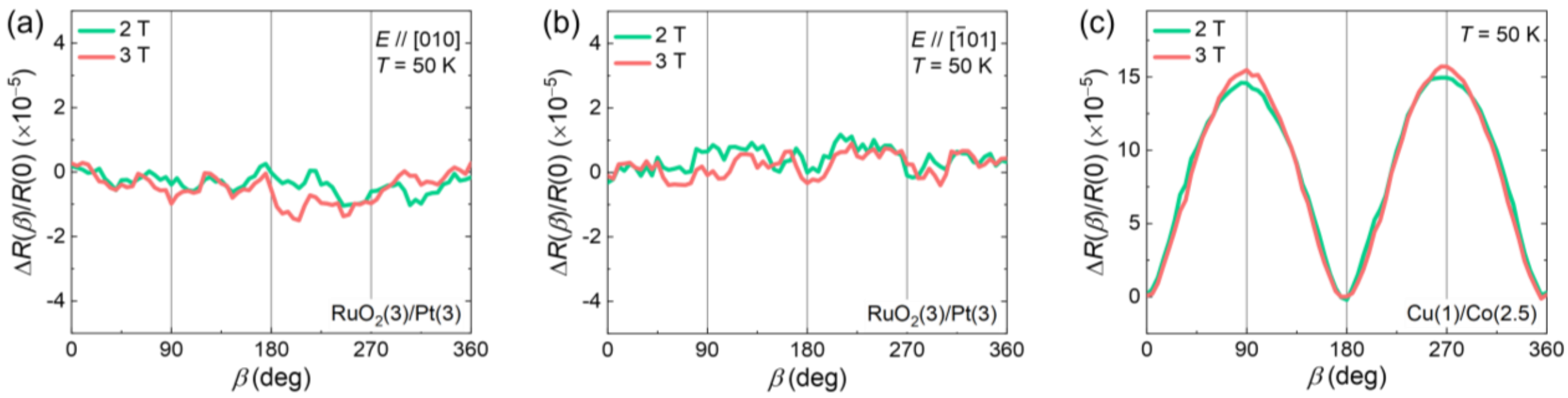

**Figure 3.** (a, b) $\Delta R(\beta)/R(0)$ measured at 50 K along $RuO_2$[010] and $RuO_2[\bar{1}01]$ in different magnetic fields a (101)-$RuO_2$(3 nm)/Pt(3 nm) bilayer. (c) $\Delta R(\beta)/R(0)$ measured at 50 K in different magnetic fields of a Cu(1 nm)/Co(2.5 nm) bilayer.

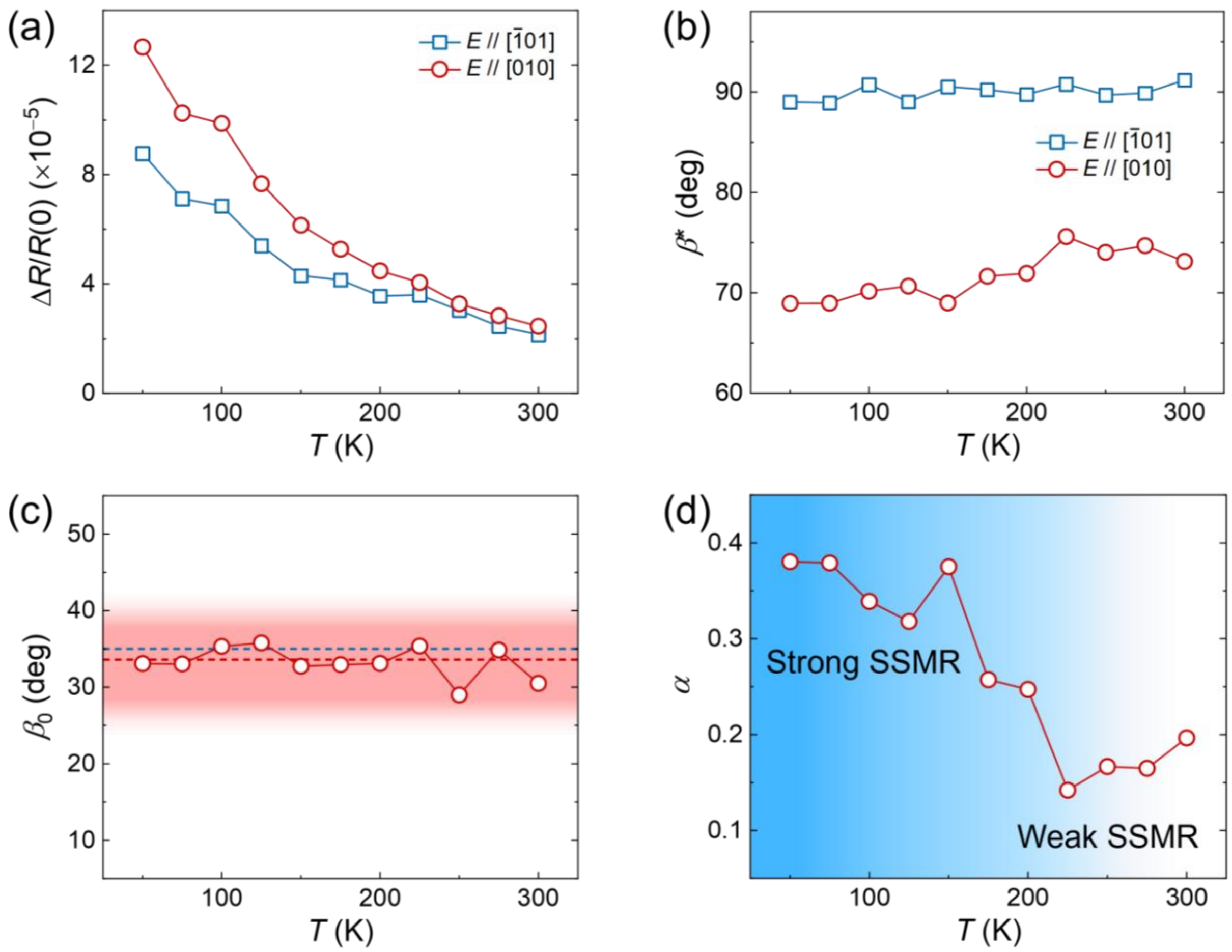

**Figure 4.** (a) Temperature ($T$)-dependent magnetoresistance ratio $\Delta R/R(0)$ and (b) phase-shift angle $\beta^*$ of the $RuO_2$(3 nm)/Co(2.5 nm) bilayer measured at 50 K along $RuO_2$[010] and $RuO_2$[$\bar{1}$01] in a magnetic field of 3 T after the subtraction of the AMR contribution. (c) Temperature-dependent out-of-plane tilting angle $\beta_0$ of the Néel vector and (d) the relative strength $\alpha$ of the pure SSMR with respect to the pure spin Hall magnetoresistance as deduced through a ternary-variable fitting process to (a) and (b). The blue dashed line in (c) denotes the out-of-plane tilting angle (~35°) of the [001]-directional Néel vector indicated by a previous x-ray scattering study.[26] The red dashed line and shaded region in (c) represent the fitted $\beta_0$ value of ~33.6° and the corresponding 95% confidence interval, respectively. The red curve denotes the $\beta_0$ calculated with Equation 14.

The table of contents entry should be 50–60 words long and should be written in the present tense. The text should be different from the abstract text.

**Spin-Splitting Magnetoresistance in Altermagnetic $RuO_2$ Thin Films**

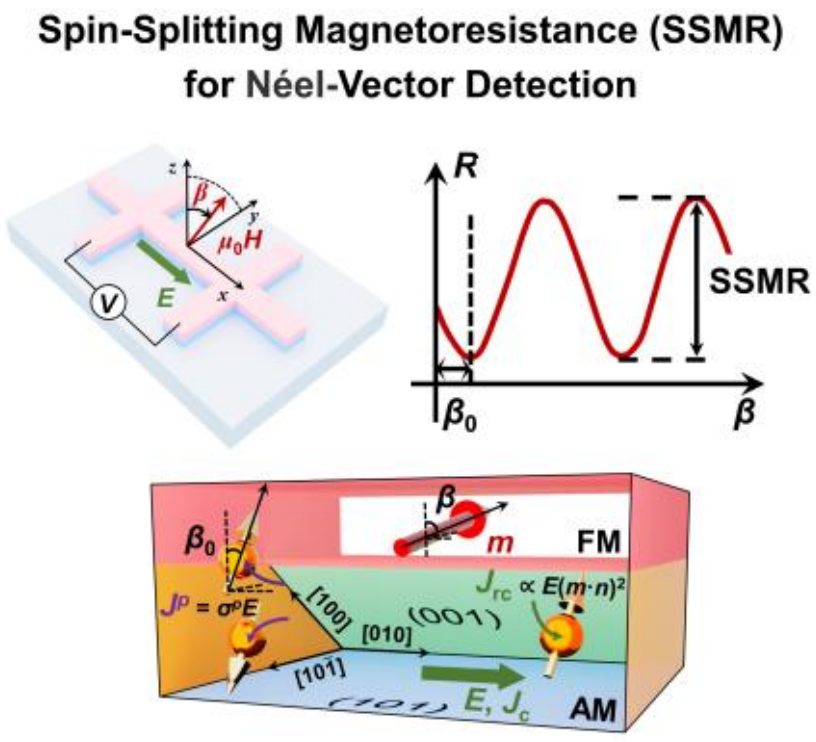

We demonstrate a novel magnetoresistance effect termed spin-splitting magnetoresistance (SSMR) in (101)-$RuO_2$/Co bilayers. The SSMR is underpinned by the spin-charge interconversion process induced by the nonrelativistic spin-splitting effect in altermagnets. Utilizing the SSMR, we reveal a [001]-oriented Néel vector in an epitaxial thin film of $RuO_2$, which evidences its altermagnetism.

# Supporting Information

**Spin-Splitting Magnetoresistance in Altermagnetic $RuO_2$ Thin Films**

*Hongyu Chen, Zian Wang, Peixin Qin*, Ziang Meng, Xiaorong Zhou, Xiaoning Wang, Li Liu, Guojian Zhao, Zhiyuan Duan, Tianli Zhang, Jinghua Liu*, Dingfu Shao*, Chengbao Jiang*, and Zhiqi Liu**

**Contents**

## Note 1 First principles calculations

First principles calculations were performed using the projector augmented-wave (PAW) method[S1] implemented in the VASP code.[S2] A plane-wave cut-off energy of 500 eV and a 16 × 16 × 16 $\vec{k}$-point mesh in the irreducible Brillouin zone were used in the calculations. The exchange and correlation effects were treated within the generalized gradient approximation (GGA) developed by Perdew-Burke-Ernzerhof (PBE).[S3] The GGA+U functional[S4,S5] with $U_{\mathrm{eff}} = 2$ eV on Ru 4*d* orbitals and $U_{\mathrm{eff}} = 5$ eV on Ti 3*d* orbitals was included in all the calculations. Spin-orbit coupling was included in all the calculations. We used the tight-binding Hamiltonians obtained from the maximally-localized Wannier functions[S6] within the Wannier90 code.[S7] The time-reversal ($\mathcal{T}$) odd and even parts of spin Hall conductivity (SHC) are given by[S8,S9]

$$\sigma_{ij}^{k} = -\frac{e\hbar}{\pi}\int \frac{d^3\vec{k}}{(2\pi)^3}\sum_{n,m}\frac{\Gamma^2\mathrm{Re}\left(\langle n\vec{k}|J_i^k|m\vec{k}\rangle\langle m\vec{k}|v_j|n\vec{k}\rangle\right)}{\left[\left(E_F - E_{n\vec{k}}\right)^2 + \Gamma^2\right]\left[\left(E_F - E_{m\vec{k}}\right)^2 + \Gamma^2\right]}, \tag{S1}$$

and

$$\sigma_{ij}^{k} = -\frac{2e}{\hbar}\int \frac{d^3\vec{k}}{(2\pi)^3}\sum_{n' \neq n}\frac{\mathrm{Im}\left(\langle n\vec{k}|J_i^k|n'\vec{k}\rangle\langle n'\vec{k}|v_j|n\vec{k}\rangle\right)}{\left(E_{n\vec{k}} - E_{n'\vec{k}}\right)^2}, \tag{S2}$$

where $J_i^k = \frac{1}{2}\{v_i, s_k\}$ is the spin-current operator, $f_{n\vec{k}}$ is the Fermi-Dirac distribution function for band $n$ and wave vector $\vec{k}$, $v_i$ and $s_k$ are velocity and spin operators, respectively, and $i, j, k = x, y, z$. A 500 × 500 × 500 $k$-point mesh was used for the integral of Equation S1 and S2. When calculating the SSE-induced $\mathcal{T}$-odd spin conductivity, $\Gamma = 50$ meV and $\Gamma = 25$ meV that determines the broadening magnitude are used. When calculating the SHE induced $\mathcal{T}$-even spin conductivity, the adaptive smearing method[S10] was used.

**Note 2 Structural Properties, Exchange Bias, and Discussions on the Altermagnetism in Thin Films**

*Structural properties*.—**Figure S1** shows the full-range x-ray diffraction pattern of a thick $RuO_2$ thin film deposited onto a (101)-oriented $TiO_2$ substrate, which indicate the high epitaxy quality of the sample.

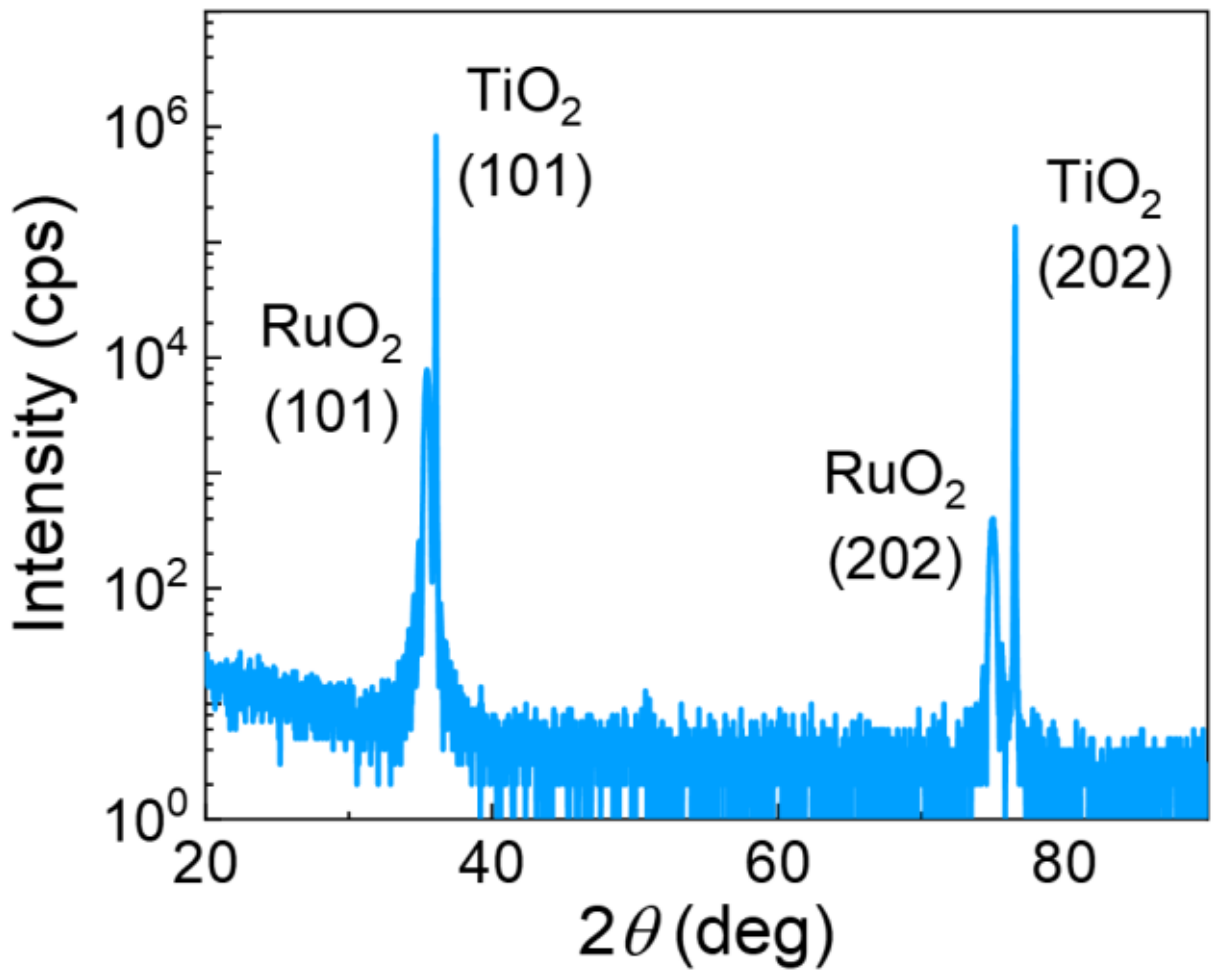

**Figure S1.** Full-range x-ray diffraction pattern of a thick (101)-$RuO_2$ thin film grown on a (101)-$TiO_2$ substrate.

*Exchange bias*.—We investigate the exchange bias (EB) of the bilayers to shed light on the antiferromagnetism of $RuO_2$. In order to establish EB, the as-deposited samples are post-annealed at 393 K in a high-vacuum VersaLab chamber for 0.5 h with a magnetic field of –1.5 T applied along $RuO_2[\bar{1}01]$. The magnetic properties of the bilayers are examined along $RuO_2[\bar{1}01]$ after field cooling down to 50 K. Furthermore, in order to highlight the role of $RuO_2$ in the EB, we prepare a 5-nm-thick Co layer capped with Al as a control sample and perform the same measurements.

**Figure S2**(a) depicts the magnetic hysteresis loop of a $RuO_2$(5 nm)/Co(5 nm) heterostructure. With the attachment of a $RuO_2$ layer, the coercive field $\mu_0 H_C$ of the Co layer is enhanced by ~14 mT and an exchange-bias field $\mu_0 H_{EB}$ of ~+0.8 mT can be extracted. We note that the $\mu_0 H_{EB}$ of an antiferromagnet/ferromagnet bilayer typically increases with the thickness of the antiferromagnet in the ultrathin limit.[S11] Therefore, the observed small $\mu_0 H_{EB}$ is well consistent with this feature and also with other previous reports.[S12,S13]

We next conduct additional experiments to further confirm the intrinsic nature of the EB. We first prepare a control sample with the same structure but perform post annealing and field cooling in an opposite magnetic field of +1.5 T. As displayed in Figure S2(b), an increased $\mu_0 H_C$ of Co comparable to that in Figure S2(a) is obtained. Moreover, the $\mu_0 H_{EB}$ changes its sign to ~–0.96 mT due to the reversed direction of the applied magnetic field in the annealing and cooling process, in good accordance with the characteristics of the EB.[S11] We also examine the magnetic properties of the bilayers with thicker $RuO_2$. As shown in Figure S2(c), the $\mu_0 H_{EB}$ is enhanced to ~+2.08 mT for the sample with a 10-nm-thick $RuO_2$ layer. Remarkably, for a $RuO_2$ thickness of 20 nm, the $\mu_0 H_C$ of Co is increased by a factor of ~2 with respect to the single layer and the $\mu_0 H_{EB}$ becomes sizable (~+2.62 mT) as displayed in Figure 2(e) in the main text. These results unambiguously demonstrate the antiferromagnetism of the $RuO_2$ thin films.

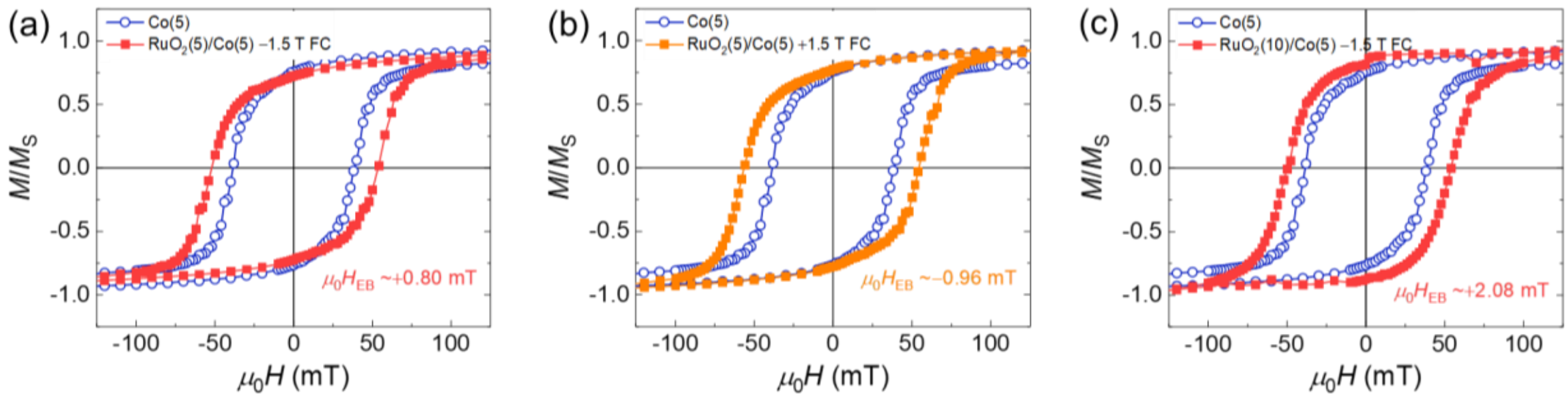

**Figure S2.** (a–c) Normalized magnetization $M/M_S$ as a function of external magnetic field $\mu_0 H$ at 50 K of a single Co layer and $RuO_2$/Co bilayers. In order to establish exchange bias, all the samples are post-annealed at 393 K for 0.5 h with a magnetic field of ±1.5 T applied along $RuO_2[\bar{1}01]$. The measurements are conducted after field cooling (FC) in the same magnetic field. The numbers in the brackets denote the thickness of the corresponding layer in units of nm.

Considering that recent reports have indicated a paramagnetic ground state mostly for bulk $RuO_2$, the emergence of antiferromagnetic order in our epitaxial thin films can be interpretated in terms of the following aspects.

(1) Epitaxial strain

Compared to bulks, epitaxial thin films are subjected to in-plane epitaxial strain. The 3-nm-thick epitaxial (101)-$RuO_2$ thin films in our study were grown on the (101) plane of $TiO_2$ single crystals with lattice constants of ~4.594 Å along [010] and ~5.464 Å along $[\bar{1}01]$. On the other hand, the in-(101)-plane lattice constants of bulk rutile-structured $RuO_2$ are ~4.492 Å along [010] and ~5.461 Å along $[\bar{1}01]$. Therefore, the lattice mismatch between $TiO_2$ and $RuO_2$ is sizable (~2.3%) along the [010] direction. For such large lattice mismatch, it typically takes a thin-film lattice 20 nm or more in thickness to fully relax the epitaxial strain. For instance, it has been shown that even a 17-nm-thick (110)-$RuO_2$ sample remains affected by epitaxial strain.[S14]

Therefore, the ultrathin 3-nm-thick sample in our study should be fully strained, as is the case in reference [S14]. The in-plane lattice constants of the sample strictly follow those of the $TiO_2$ substrate, and the sample is subjected to an in-plane strain of ~+2.3% along [010]. Moreover, the out-of-plane lattice constant of our sample should also be largely different from that of bulk, as the out-of-plane interplanar spacing of the 25-nm-thick sample deduced from the diffraction peak at $2\theta$ ~74.974° in Figure S1 remains ~0.9% smaller than that of a bulk.

Lastly, we point out that the physical properties of $RuO_2$ tend to be sensitive to strain. For instance, it has been theoretically suggested that altermagnetism can be introduced to $RuO_2$ by strain[S15]. In addition, recent studies have demonstrated strain-stabilized superconductivity in epitaxial $RuO_2$ thin films[S16,S17]. Therefore, we expect that the epitaxial strain may result in the antiferromagnetic order in our samples.

(2) Surface/interface symmetry breaking

Distinct from bulks, the macroscopic properties of ultrathin films can be greatly affected by symmetry breaking at the surface/interface wherein electrons are exposed to asymmetric potential fields. We have recently started a study to investigate the surface/interface symmetry breaking and the altermagnetism in epitaxial $RuO_2$ thin films. We have indeed found some meaningful results, which will be discussed in detail in our upcoming work.

(3) Charge depletion at the surface/interface

As shown in Figure S7(b), $RuO_2$ is a conductive oxide exhibiting low electrical resistivity and metallic transport behavior, while $TiO_2$ is a wide bandgap semiconductor. Therefore, we expect the contact of conductive $RuO_2$ and insulating $TiO_2$/atmosphere would lead to significant charge depletion at the surface/interface, thereby prominently affecting the properties of our ultrathin samples.

## Note 3 Exclusion of Experimental Artifacts of the Phase Shift

We can rule out the following possible origins of the phase-shifted angular dependence.

(1) Agular Offset in the Measurement

This cannot lead to the observed phase shift since there is no anomaly in the $\beta$-dependent magnetoresistance measured along $RuO_2[\bar{1}01]$ with extrema occurring at 0°, 90°, 180°… in Figure 2(e) in the main text.

(2) Unsaturated Magnetization of Co

It has been shown that ferromagnets with strong magnetic anisotropy can exhibit distorted $R(\beta)$ curves similar to Figure 2(e) in the main text when the applied magnetic field is too weak to completely align the magnetic moments.[S18] To rule out this possibility, we first characterize the magnetic anisotropy of the Co layer in a (101)-$RuO_2$(3 nm)/Co(2.5 nm) sample. As depicted in **Figure S3**(a,b), the as-deposited thin film show strong in-plane anisotropy with an anisotropy field lower than 1 T from 50 to 300 K. No observable anomaly can be found in both the in-plane and the out-of-plane magnetic hysteresis loops. Furthermore, the magnetic properties of the Co layer are almost unchanged after micro-nano processing, as evidenced by the high similarity between the out-of-plane magnetization-magnetic field curves of the as-deposited thin film in Figure S3(b) and the Hall resistance-magnetic field curves of the patterned Hall bar in Figure S3(c). Therefore, a magnetic field of 3 T is sufficient to align the magnetization of the Co layer.

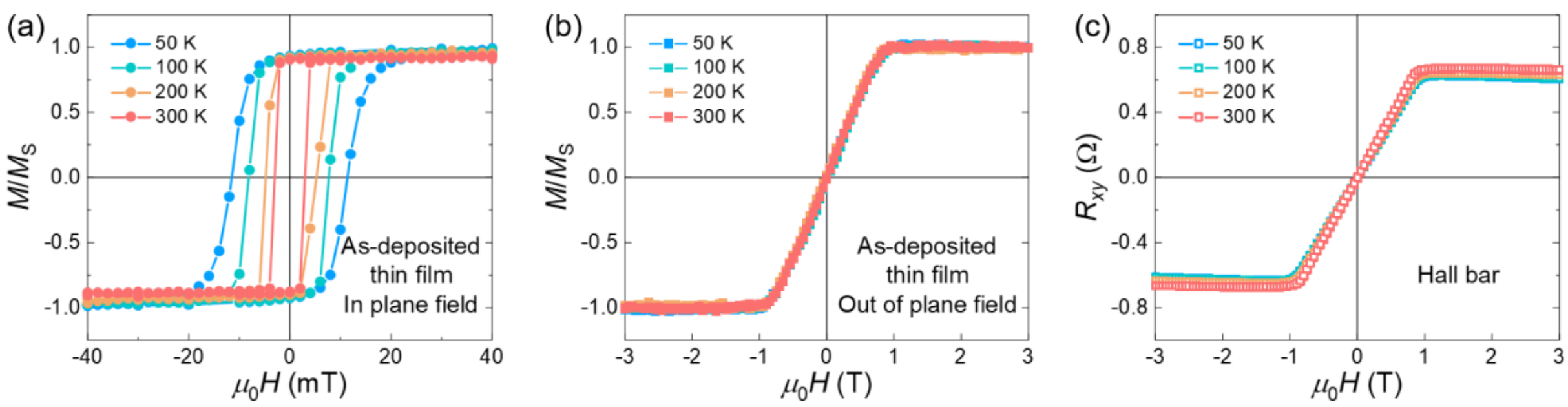

**Figure S3.** (a) In-plane and (b) out-of-plane $M/M_S$ as a function of $\mu_0H$ at various temperatures for an as-deposited (101)-$RuO_2$(3 nm)/Co(2.5 nm) thin film. (c) Anomalous Hall effect of the sample at different temperatures after micro-nano processing.

To further exclude this possibility, rotation-polarity- and magnetic field-dependent $R(\beta)$ are studied. It is found that the $R(\beta)$ curves measured with $\beta$ rotated from 0° to 360° and from 360° back to 0° overlap well with each other [**Figure S4**(a)], in sharp contrast to the hysteresis behavior in ref. [S18]. Moreover, the $\Delta R(\beta)/R(0)$ curves are almost unchanged for magnetic fields larger than 1.2 T [Figure S4(b)], solidifying that the anomalous $\Delta R(\beta)/R(0)$ is a robust and intrinsic phenomenon.

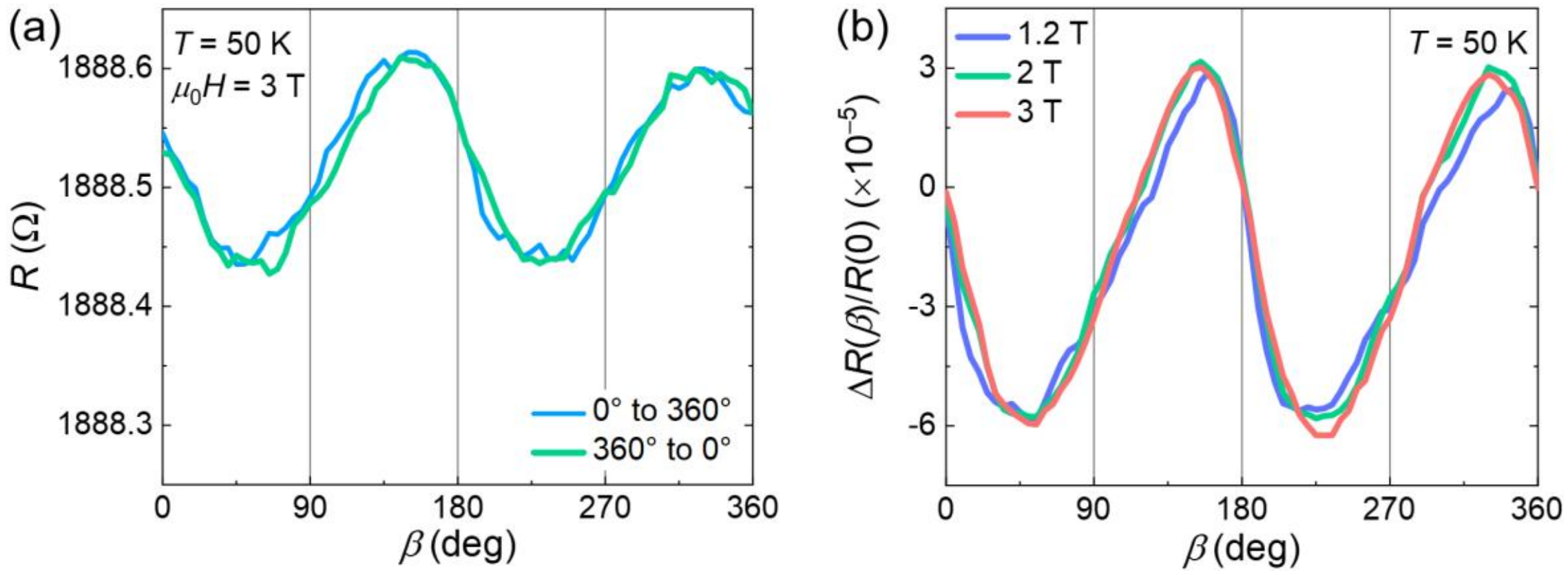

**Figure S4.** (a) $\beta$-dependent longitudinal resistance $R$ of the (101)-$RuO_2$(3 nm)/Co(2.5 nm) bilayer measured at 50 K along $RuO_2$[010] in a magnetic field of 3 T with back-and-forth sample rotation. (b) $\beta$-dependent magnetoresistance $\Delta R(\beta)/R(0)$ of the sample measured at 50 K along $RuO_2$[010] in different magnetic fields.

(3) Anisotropic Magnetoresistance of $RuO_2$

Recent studies have unveiled that (101)-oriented rutile-structured $IrO_2$ thin films without long-range magnetic order can exhibit abnormal angular-dependent planar Hall resistance with a phase-shift feature due to the low symmetry.[S19] In order to ascertain whether this nonmagnetic mechanism can lead to the unusual $\Delta R(\beta)/R(0)$ in Figure 2(e) in the main text, we construct a parallel circuit model as illustrated in **Figure S5**(a) to disentangle the contribution from Co and $RuO_2$ to the measured magnetoresistance. The longitudinal resistances of the Co and the $RuO_2$ layer at $\beta = 0°$ are denoted as $R_1$ and $R_2$, respectively. According to this model, the longitudinal resistance of the bilayer at $\beta = 0°$ is

$$R_L(0) = \frac{R_1 R_2}{R_1 + R_2} \tag{S3}$$

With the sample rotated by an angle $\beta$, $R_1$ and $R_2$ change by $\Delta R_1(\beta)$ and $\Delta R_2(\beta)$, respectively, and the longitudinal resistance of the bilayer becomes

$$R_L(\beta) = \frac{(R_1 + \Delta R_1)(R_2 + \Delta R_2)}{R_1 + \Delta R_1 + R_2 + \Delta R_2} \tag{S4}$$

Therefore, the measured magnetoresistance is

$$\Delta R(\beta)/R(0) = \frac{R_L(\beta) - R_L(0)}{R_L(0)}$$

$$= \frac{\Delta R_2 R_1}{R_2(R_1 + \Delta R_1 + R_2 + \Delta R_2)} + \frac{\Delta R_1 R_2}{R_1(R_1 + \Delta R_1 + R_2 + \Delta R_2)} + \frac{\Delta R_1}{R_1}\frac{\Delta R_2}{R_2}\frac{R_1+R_2}{R_1 + \Delta R_1 + R_2 + \Delta R_2}$$

$$\approx \frac{\Delta R_2}{R_2}\frac{R_1}{R_1+R_2} + \frac{\Delta R_1}{R_1}\frac{R_2}{R_1+R_2} \quad \text{(S5)}$$

Since $\Delta R_2(\beta)$ contains contributions from the magnetoresistance of $RuO_2$ ($\Delta R_{21}$) and the bilayer ($\Delta R_{22}$), Equation S5 can be rewritten as

$$\Delta R(\beta)/R(0) \approx \frac{\Delta R_{21}}{R_2}\frac{R_1}{R_1+R_2} + \frac{\Delta R_{22}}{R_2}\frac{R_1}{R_1+R_2} + \frac{\Delta R_1}{R_1}\frac{R_2}{R_1+R_2} \quad \text{(S6)}$$

And the contribution from the magnetoresistance of $RuO_2$ to $\Delta R(\beta)/R(0)$ is $\frac{\Delta R_{22}}{R_2}\frac{R_1}{R_1+R_2}$.

We thus measure the $\beta$-dependent magnetoresistance of a 3-nm-thick $RuO_2$ layer to determine $\frac{\Delta R_{22}}{R_2}$. As shown in Figure S5(b, c), $\frac{\Delta R_{22}}{R_2}$ is less than $10^{-5}$. On the other hand, the $\Delta R(\beta)/R(0)$ of the bilayer can reach a value of ~$9 \times 10^{-5}$. Therefore, the contribution from the anisotropic magnetoresistance of $RuO_2$ to the measured $\Delta R(\beta)/R(0)$ is negligible and cannot lead to the remarkable phase shift in Figure 2(e) in the main text.

Consequently, we conclude that the phase-shifted angular-dependent magnetoresistance is an intrinsic property of the (101)-$RuO_2$(3 nm)/Co(2.5 nm) bilayer.

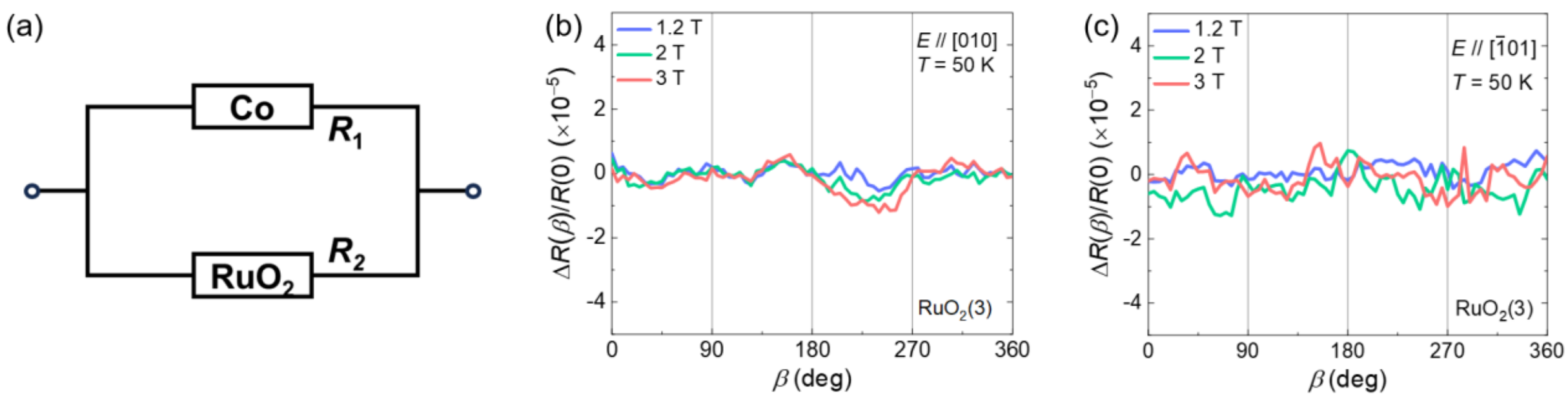

**Figure S5.** (a) Parallel circuit model of the bilayers. (b, c) $\Delta R(\beta)/R(0)$ measured along the [010] and the [$\bar{1}$01] direction at 50 K and in different magnetic fields of a 3-nm-thick $RuO_2$ thin film.

## S4 Discussion on the Magnon-Excitation-Induced Magnetoresistance

It has been proposed that the magnon excitations in a nonmagnet/ferromagnet bilayer induced by an electric field can also contribute to the SMR.[S20] Since magnon population is sensitive to temperature and magnetic fields, we study the field-dependent magnetoresistance of the sample at 50 and 300 K to examine the role of magnons. We employ the difference between the sample resistances measured in an magnetic field along the in-plane $y$ direction and the out-of-plane $z$ direction, *i.e.*, $R_y - R_z$, to capture the field-dependence of $\Delta R/R(0)$. As displayed in **Figure S6**, the $R_y - R_z$ measured along $RuO_2$[010] at both 50 and 300 K are almost a constant in magnetic fields larger than 1 T. Note that the $\Delta R(\beta)/R(0)$ curves at 50 K in different magnetic fields are also almost identical [Figure S4(b)]. This demonstrates that magnon excitations play a negligible role in generating the $\Delta R(\beta)/R(0)$.

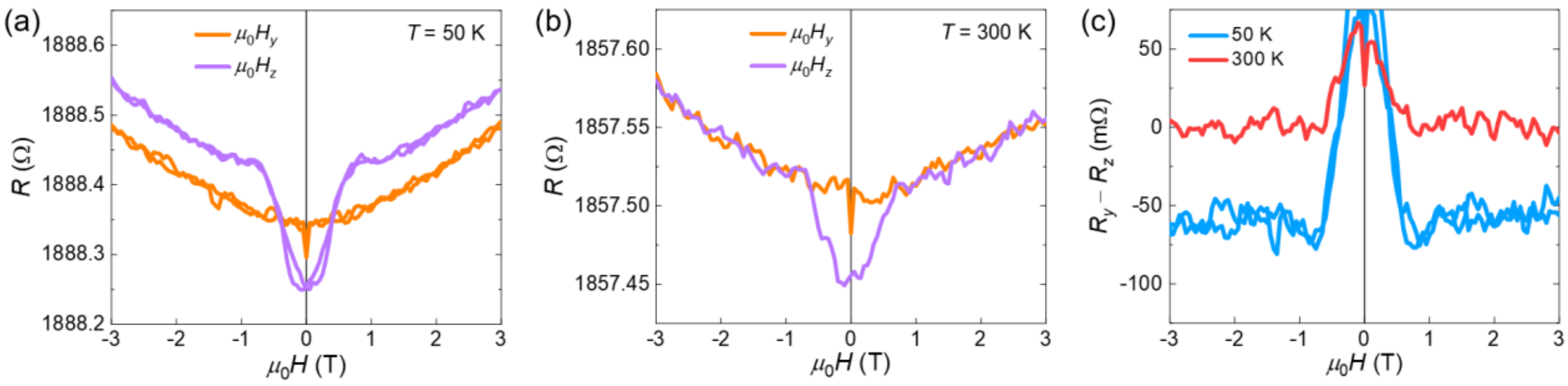

**Figure S6.** (a, b) Field-dependent $R$ of the (101)-$RuO_2$(3 nm)/Co(2.5 nm) bilayer measured along $RuO_2$[010] at 50 and 300 K. The magnetic field is applied along the in-plane $y$ direction (perpendicular to the current) and the out-of-plane $z$ direction. (c) Field-dependent difference between the $R$ in the $y$- and $z$-direction magnetic field $R_y - R_z$ at 50 and 300 K.

## S5 Subtraction of the Anisotropic Magnetoresistance of Co

In the main text, we find that the ferromagnetic AMR of Co is comparable to the overall magnetoresistance of the bilayer. We thus try to subtract this AMR contribution from the measured data. According to Equation S5, the AMR of Co, *i.e.*, $\frac{\Delta R_1}{R_1}$, participate in the $\Delta R(\beta)/R(0)$ of the bilayer in the form of $\frac{\Delta R_1}{R_1}\frac{R_2}{R_1+R_2}$. Therefore, we first determine the intrinsic resistivity of $RuO_2$ and Co in the bilayers according to the parallel circuit model shown in Figure S5(a). The conductance of $RuO_2(t_{RO})$/Co(2.5 nm) bilayers with varied thickness of $RuO_2$ can be written as

$$\frac{1}{R_L} = \frac{t_{Co}}{\rho_{Co}}\frac{W}{L} + \frac{t_{RO}}{\rho_{RO}}\frac{W}{L} \tag{S7}$$

wherein $L$ and $W$ are distance between the two longitudinal voltage probe and the channel width of the Hall bar, respectively; $t_{RO}$ and $\rho_{RO}$ is the thickness and the resistivity of $RuO_2$, respectively; $t_{Co}$ and $\rho_{Co}$ are the thickness and resistivity of Co, respectively. Therefore, we have

$$\frac{1}{R_L}\frac{L}{W} = \frac{t_{Co}}{\rho_{Co}} + \frac{t_{RO}}{\rho_{RO}} \tag{S8}$$

Here $\frac{1}{R_L}\frac{L}{W}$ is the sheet conductance of the bilayers. Note that the measured $\frac{L}{W}$ values of our samples are in the range of 4–7. According to Equation S8, $\rho_{Co}$ and $\rho_{RO}$ can be extracted from a linear fitting of the relation between $\frac{1}{R_L}\frac{L}{W}$ and $t_{RO}$.[S21] As shown in **Figure S7**(a), $\frac{1}{R_L}\frac{L}{W}$ is indeed proportional to $t_{RO}$, indicating that the $\rho_{Co}$ and $\rho_{RO}$ are almost unchanged in the $RuO_2(t_{RO})$/Co(2.5 nm) bilayers. We find that the difference in the resistance measured along $RuO_2$[010] and $RuO_2[\bar{1}01]$ is negligible, and the deduced $\rho_{RO}$ and $\rho_{Co}$ are displayed in Figure S7(b,c). Note that the $\rho_{RO}$ on the order of $10^2$ μΩ cm is well consistent with other previous studies.[S14,S22]

We then measure the AMR of a 2.5-nm-thick Co layer, as depicted in **Figure S8**(a). The contribution from the Co layer to the overall magnetoresistance of the bilayers can thus be subtracted according to Equation S5. For instance, the magnetoresistance solely induced by the nonequilibrium spin distribution in the $RuO_2$(3 nm)/Co(2.5 nm) bilayer at 50 K along $RuO_2$[010] is displayed in Figure S8(b). We also subtract the AMR contribution from the $RuO_2$(3 nm)/Cu(1 nm)/Co(2.5 nm) sample according to the AMR of Cu(1 nm)/Co(2.5 nm) and its resistivity of ~278 μΩ cm as shown in Figure S8(c).

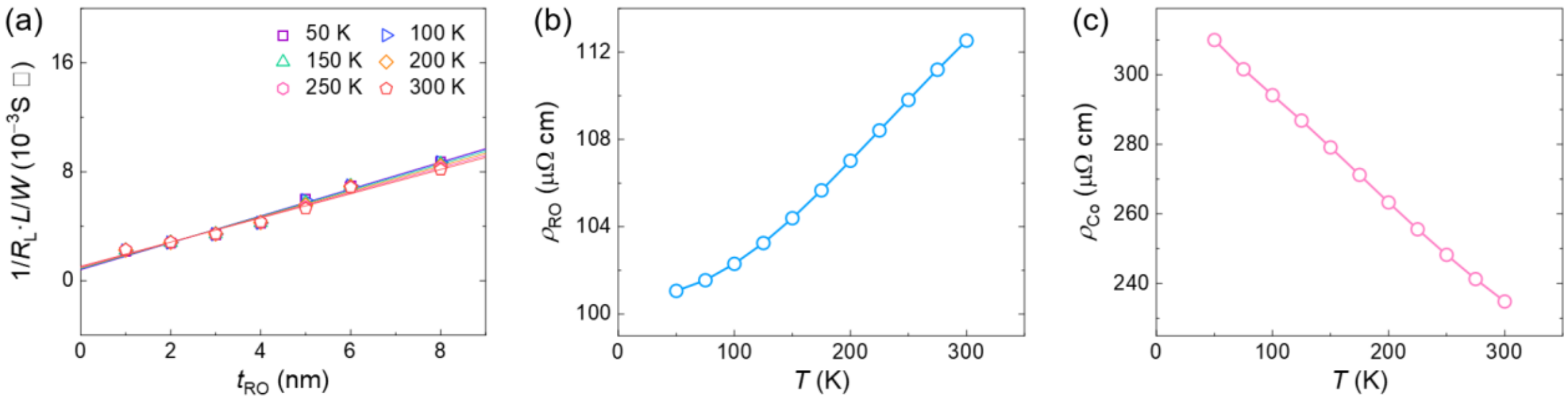

**Figure S7.** (a) Linear fitting of the relation between the sheet conductance of the $RuO_2(t_{RO})$/Co(2.5 nm) bilayers and the thickness $t_{RO}$ of the $RuO_2$ layer at various temperatures. (b) Resistivity $\rho_{RO}$ of $RuO_2$ and (c) resistivity $\rho_{Co}$ of Co at various temperatures deduced from (a).

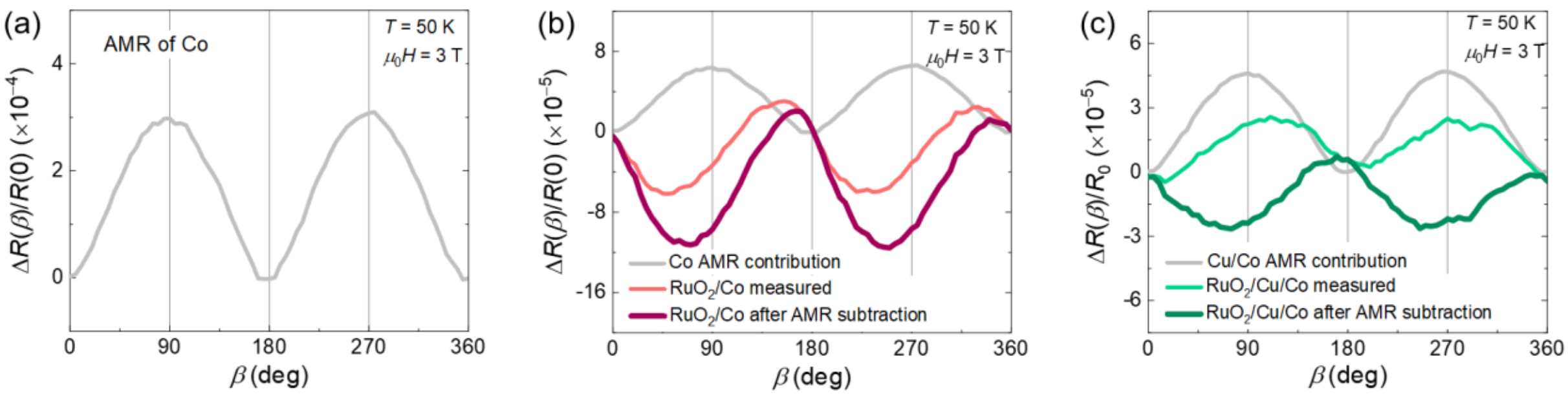

**Figure S8.** (a) Anisotropic magnetoresistance (AMR) measured at 50 K in a magnetic field of 3 T for a 2.5-nm-thick Co layer. (b, c) $\Delta R(\beta)/R(0)$ of the $RuO_2$(3 nm)/Co(2.5 nm) bilayer and the $RuO_2$(3 nm)/Cu(1 nm)/Co(2.5 nm) trilayer measured at 50 K along $RuO_2$[010] in a magnetic field of 3 T before and after the subtraction of the AMR of Co.

## S6 Discussion on the Spin Hall Effect and the Spin-Splitting Effect of (101)-$RuO_2$

As illustrated in **Figure S9**, a (101)-oriented thin film with a rutile structure possesses only one glide mirror plane $M_{010}$ normal to the film. Therefore, for the out-of-plane spin current induced by an electric field (parallel to the *x* axis of the coordinate system) along [$\bar{1}$01] (parallel to $M_{010}$), only conventional in-plane *y*-components of spin polarization is allowed. On the other hand, for the out-of-plane spin current induced by an electric field along [010] (perpendicular to $M_{010}$), unconventional out-of-plane *z*-components of spin polarization is additionally permitted. Note that such symmetry analyses are not confined to a specific mechanism for spin-current generation. Therefore, in principle, both the nonrelativistic spin-splitting effect (SSE) and the relativistic spin Hall effect (SHE) can lead to the anisotropic behavior of $\Delta R(\beta)/R(0)$ in Figure 2(e) in the main text.

We first concentrate on the SHE of $RuO_2$, which exists in both altermagnetic and nonmagnetic $RuO_2$. We calculate the SHE, *i.e.*, the time-reversal ($\mathcal{T}$)-even components of the spin conductivity tensor, for altermagnetic $RuO_2$ (**Table S1**). We find that the *z*-components of spin polarization engendered by the SHE are subtle and cannot lead to the remarkable phase shift. In addition, recent studies have systematically characterized the SHE in (101)-oriented epitaxial thin films of $IrO_2$, a rutile-structured material without long-range magnetic order.[S23,S24] It is found that the unconventional *z*-components of spin polarization generated by an electric field along [010] is negligibly small (almost zero) despite its theoretical allowance. This further indicates the incapability of the SHE in producing sizable unconventional spin currents and thus the remarkable phase shift in $\Delta R(\beta)/R(0)$ in such thin films.

We next calculate the SSE, *i.e*, the $\mathcal{T}$-odd components of the spin conductivity tensor, of altermagnetic $RuO_2$. As displayed in Table S1, for an electric field applied along [010], the SSE is much larger than the SHE. Remarkably, the out-of-plane tilting angle of the spin polarization of the spin current generated by the SSE can be calculated as $\cot^{-1}(\sigma^z/\sigma^y) \approx 35°$, parallel to that of the Néel vector of $RuO_2$. On the other hand, for an electric field applied along [$\bar{1}$01], the SSE only engenders out-of-plane spin current with *y*-components of spin polarization, and is comparable to or even smaller than the SHE.

It is worth noting that the SSE can be strongly disturbed by electron scattering because it is a property closely related to the Fermi surface.[S23,S25–S27] We highlight this feature in our calculations utilizing different broadening of the Γ function to mimic the effect of temperature (electron scattering) on the SSE. It is shown that the raise of temperature (Γ function broadening) remarkably decreases the SSE due to the enhanced electron scattering. In contrast, the SHE has been shown to be less susceptible to temperature.[S28,S29]

We finally discuss the competition between the SHE and the SSE in altermagnetic $RuO_2$. Although Table S1 indicates a one-order-of-magnitude larger SSE than the SHE at low temperature, we note that this can only be true for a single-domain $RuO_2$. The epitaxial $RuO_2$ thin films in this study is presumably in a multidomain state with the Néel vector aligned along either [001] or [00$\bar{1}$] in each domain. Consequently, the SSE of the magnetic domains with antiparallel Néel vectors would compensate each other due to its $\mathcal{T}$-odd nature, leading to a net SSE smaller than the calculation results. This is corroborated by the deduced $\beta^*$ of ~69° after subtracting the AMR of Co in Figure 2(f) in the main text. According to the (S)SMR scenario, $\beta^*$ describes the out-of-plane tilting angle of the spin polarization of the engendered spin current. If the SSE dominates the spin-current generation, a $\beta^*$ ~35° would be expected. A much larger $\beta^*$ indicates that the spin polarization contains more $y$-components than that can be produced by the SSE, highlighting the existence of the SHE that is comparable to the SSE in magnitude.

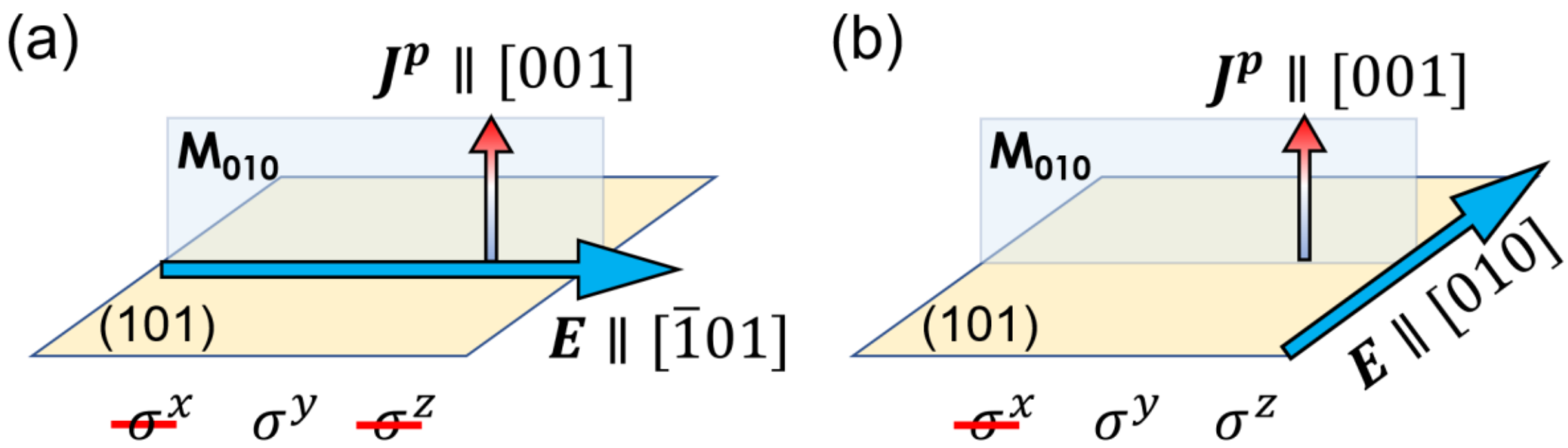

**Figure S9.** (a, b) Schematic illustration of the out-of-plane spin current $\boldsymbol{J}^{\mathbf{p}}$ induced by an electric field $E$ (parallel to the $x$ axis of the coordinate system) in (101)-$RuO_2$ thin films. An electric field $E$ generates a $\boldsymbol{J}^{\mathbf{p}}$ as $J^p = \sigma^p E$, where $\boldsymbol{\sigma}^{\boldsymbol{p}} = (\sigma^x, \sigma^y, \sigma^z)$ is the spin conductivity and $\boldsymbol{p}$ is the spin polarization direction. For $E$ // [$\bar{1}$01], only the spin conductivity component $\sigma^y$ is allowed by the glide mirror plane $M_{010}$. For $E$ // [010], the spin conductivity component $\sigma^x$ is forbidden by $M_{010}$.

**Table S1.** Calculated spin conductivity for $RuO_2$ (101) film in units of $(\frac{\hbar}{2e})$ $(\Omega\ \mathrm{cm})^{-1}$, where $\sigma_{zx}$ denotes spin conductivity for $E$ along $x$ ([010]) direction, and $\sigma_{zy}$ denotes spin conductivity for $E$ along $y$ ($[\bar{1}01]$) direction.

| | $\sigma_{zx}^{x}$ | $\sigma_{zx}^{y}$ | $\sigma_{zx}^{z}$ | $\sigma_{zy}^{x}$ | $\sigma_{zy}^{y}$ | $\sigma_{zy}^{z}$ |
|---|---|---|---|---|---|---|
| $\mathcal{T}$-odd $\Gamma \approx 25$ meV | 0 | 1237.7 | 1728.6 | 297.7 | 0 | 0 |
| $\mathcal{T}$-odd $\Gamma \approx 50$ meV | 0 | 627.4 | 888.2 | -147.2 | 0 | 0 |
| $\mathcal{T}$-even | 0 | 175.6 | -33 | -256.2 | 0 | 0 |

**Note 7 Discussion on the Effectiveness of the Fitting Model**

In the main text, we develop a method to disentangle the SSMR and the SMR via a ternary-variable fitting process. Here we discuss its effectiveness in terms of two aspects.

(1) Applicability of the method

We applied some approximations to derive the fitting function Equation in our method, aiming to both minimize the number of necessary fitting parameters and enhance the fitting efficiency. Therefore, the applicability and effectiveness of these approximations coincides with that of the proposed model. We note that the approximation pertaining to the prominent and anisotropic SSE of (101)-$RuO_2$, *i.e.*, a much stronger SSE-generated out-of-plane spin current for a current along [010] than along [$\bar{1}$01], is important in our study, but may lack generalizability to other material systems. In other words, for systems with weak and (or) nearly-isotropic SSE, additional fitting parameters shall be incorporated in the fitting function, which may impact the effectiveness of the method. Therefore, we emphasize that our method could not be universal and its effectiveness might depend on specific material systems.

(2) Effectiveness of the fitting process

In a specific case, the effectiveness of the fitting process can be evaluated based on the fitting results and regression statistics. In our study, the ternary-variable fitting yields a $\beta_0$ (the phase shift of pure SSMR as defined in the main text) of ~33.6° with a stand error of ~3.3° and a 95% confidence interval of [26.0°, 41.2°] and an $\eta$ (a parameter describing the anisotropy of the spin transport in $RuO_2$ as defined in the main text) of 0.84 with a stand error of ~0.03 and a 95% confidence interval of [0.77, 0.91]. The deduced $\beta_0$ and $\eta$ values are reasonable in physics, and the standard errors and 95% confidence intervals are acceptable. Therefore, the $\alpha$ (the relative strength of the pure SSMR with respect to the pure SMR as defined in the main text) calculated with $\beta_0$ and $\eta$ is reliable. The adjusted $R$ squared of the fitting is ~0.992, indicating that the measured temperature dependence of $\Delta R/R(0)$ and $\beta^*$ in Figure 4(a, b) can be well captured by the proposed model. In order to straightforwardly display the fitting result of $\beta_0$ and the effectiveness of our model, we plot the fitted $\beta_0$ (red dashed line) with its 95% confidence interval (red-shaded region) in Figure 4(c) in the main text. We also calculate $\beta_0$ with the fitted $\eta$ (red curve). We find that the calculated $\beta_0$ exhibits stochastic fluctuations within narrow bounds around the fitted value with no substantial deviation, thereby confirming the soundness of our approach.